\documentclass[aps,twocolumn]{revtex4}
\usepackage{bm}
\usepackage{amsmath}
\usepackage{graphicx}
\usepackage{subfigure}
\usepackage[usenames,dvipsnames]{color}
\definecolor{darkblue}{RGB}{0,0,196}
\usepackage[colorlinks=true,linkcolor=darkblue,citecolor=darkblue,urlcolor=darkblue]{hyperref}
\usepackage{setspace}
\usepackage{hyperref}
\usepackage{xcolor}
\hypersetup{
  colorlinks   = true, 
  urlcolor     = red, 
  linkcolor    = blue, 
  citecolor   = blue 
}
\usepackage{footmisc}
\usepackage[makeroom]{cancel}
\usepackage{comment}
\def\be{\begin{equation}}
\def\ee{\end{equation}}
\def\ba{\begin{eqnarray}}
\def\ea{\end{eqnarray}}
\usepackage{graphicx}
\usepackage{amsmath,bbm}
\usepackage{amssymb,bm}
\usepackage{yfonts}

\begin{document}
\title{Estimation of Impact Parameter and Transverse Spherocity in heavy-ion collisions at the LHC energies using Machine Learning}
\author{Neelkamal Mallick$^{1}$}
\author{Sushanta Tripathy$^{3}$}
\author{Aditya Nath Mishra$^{4}$}
\author{Suman Deb$^{1}$}
\author{Raghunath Sahoo$^{1,2,}$\footnote{Corresponding author: Raghunath.Sahoo@cern.ch }}
\affiliation{$^{1}$Department of Physics, Indian Institute of Technology Indore, Simrol, Indore 453552, India}
\affiliation{$^{2}$CERN, CH 1211, Geneva 23, Switzerland}
\affiliation{$^{3}$INFN - sezione di Bologna, via Irnerio 46, 40126 Bologna BO, Italy}
\affiliation{$^{4}$Wigner Research Center for Physics, H-1121 Budapest, Hungary}

\begin{abstract}
\noindent
Recently, machine learning (ML) techniques have led to a range of numerous developments in the field of nuclear and high-energy physics. In heavy-ion collisions, the impact parameter of a collision is one of the crucial observables which has a significant impact on the final state particle production. However, calculation of such a quantity is nearly impossible in experiments as the length scale ranges in the level of a few fermi. In this work, we implement the ML-based regression technique via Boosted Decision Trees (BDTs) to obtain a prediction of impact parameter in Pb-Pb collisions at $\sqrt{s_{\rm NN}}$ = 5.02 TeV using A Multi-Phase Transport (AMPT) model. In addition, we predict an event shape observable, transverse spherocity in Pb-Pb collisions at $\sqrt{s_{\rm NN}}$ = 2.76 and 5.02 TeV using AMPT and PYTHIA8 based on Angantyr model. After a successful implementation in small collision systems, the use of transverse spherocity in heavy-ion collisions has potential to reveal new results from heavy-ion collisions where the production of a QGP medium is already established. We predict the centrality dependent spherocity distributions from the training of minimum bias simulated data and it was found that the predictions from BDTs based ML technique match with true simulated data. In the absence of experimental measurements, we propose to implement Machine learning based regression technique to obtain transverse spherocity from the known final state observables in heavy-ion collisions. 

\pacs{}
\end{abstract}
\date{\today}
\maketitle 

\section{Introduction}
\label{intro}
A deconfined state of quarks and gluons, also known as Quark Gluon Plasma (QGP) is believed to be produced in ultra-relativistic heavy-ion collisions at the Large Hadron Collider (LHC). However, due to its very short lifetime we do not have any direct evidence of possible QGP formation, instead several indirect signatures such as strangeness enhancement, quarkonia suppression, direct photon measurements, elliptic flow etc. suggest that formation of QGP is highly probable in such collisions~\cite{Bass:1998vz}. Such observables are usually studied as a function of centrality classes of the collisions which are determined by the impact parameter ($b$). However, obtaining the impact parameter values from experiments is still challenging as its value ranges in few femtometers ($fm$). Thus, in experiments the centrality classes are inferred from final state charged-particle multiplicities and sometimes from the transverse energy distribution. In the hindsight, it would benefit the experiments if one can successfully implement Machine Learning (ML) based technique to obtain the impact parameter in a precise way.

Historically, the results from proton-proton (pp) collisions are considered as a baseline for understanding the results obtained for heavy-ion collisions. To understand the recent measurements of heavy-ion-like behaviors~\cite{Khachatryan:2016txc,ALICE:2017jyt} in pp collisions at the LHC, a new event classifier known as transverse spherocity, an event shape observable, has been introduced.~\cite{Cuautle:2014yda,Cuautle:2015kra,Salam:2009jx,Bencedi:2018ctm,Banfi:2010xy,Khuntia:2018qox,Tripathy:2019blo,Tripathy:2020jue,Tripathy:2020lla}. After its successful implementation in small collision systems, the use of transverse spherocity in heavy-ion collisions has a potential to reveal new physics where the production of a QGP medium is already established. In our recent publication~\cite{Mallick:2020yih}, we have explicitly used transverse spherocity in heavy-ion collisions for the first time to study the final state particle correlations and azimuthal anisotropy as a function of transverse spherocity in A Multi-Phase Transport (AMPT) model. A strong anti-correlation of transverse spherocity with the ellipticity of the events in heavy-ion collisions was observed. It was found that low transverse spherocity events contribute significantly to the elliptic flow while high transverse spherocity events have nearly zero elliptic flow. 
This indicates that transverse spherocity can be used as a new event classifier in heavy-ion collisions.
However, so far no measurement has been performed in heavy-ion collisions as a function transverse spherocity in any of the LHC experiments due to the fact that such a measurement becomes computationally challenging in heavy-ion collisions. Thus, the application of ML based regression technique to obtain transverse spherocity from the known final state experimental observables would be very useful in the current scenario. 

Recently, machine learning techniques have led to a range of numerous developments in the field of high-energy physics (HEP) along with in different fields of physics~\cite{Carleo:2019ptp,Armitage:2018own,Radovic:2018dip,Albertsson:2018maf,ML-HEP-com,Ortiz:2020rwg,Ortiz:2021peu}. For several years different machine learning algorithms have been used to determine the impact parameter~\cite{Bass:1993vx,Li:2020qqn,David:1994qc,Bass:1996ez,Haddad:1996xw,DeSanctis:2009zzb}.  Thus, it is timely to implement ML based techniques to obtain the impact parameter and transverse spherocity distributions at the LHC energies. Machine learning methods are designed to exploit large datasets in order to reduce complexity. They also help to find new features in data. Currently, the most frequently used machine learning algorithms in high-energy physics are Boosted Decision Trees (BDT)~\cite{Roe:2004na} and Neural Networks (NN). Usually, machine learning model is trained for variables relevant to a particular physics problem, which can be classified into either classification or regression problem. In both the cases, training the model is the most time consuming step for both humans and computer CPU, while the inference stage is relatively inexpensive. Thus, machine learning models are gaining lots of popularity in different fields of basic sciences. BDTs and NNs are typically used to classify particles and events. However, they are also used for regression, where a continuous function is learned and gives a prediction of an observable which is usually cumbersome to obtain from real experiments. In this work, we implement ML-based regression technique via BDT to obtain predictions for impact parameter and spherocity distributions in Pb-Pb collisions at $\sqrt{s_{\rm NN}}$ = 2.76 and 5.02 TeV using A Multi-Phase Transport (AMPT)~\cite{AMPT2} and PYTHIA8 (Angantyr) model~\cite{Bierlich:2018xfw}. For machine learning, we have used a python based machine learning package, called as scikit-learn software package \cite{sklearn}. We have specifically used the {\em GradientBoostingRegressor} module inside {\em sklearn.ensemble} framework. For our study, we use final state charged-particle multiplicity and mean transverse momentum as the input variables for the predictions of impact parameter and transverse spherocity.

This paper is organised as follows. We start with a brief introduction on event generation and target observables for ML in section~\ref{section2}. Then, in section~\ref{section3} we provide a detailed procedure of ML based regression technique along with few quality assurance plots to obtain the impact parameter and transverse spherocity from heavy-ion collisions at the LHC energies using event generators such as AMPT and PYTHIA8 (Angantyr) models. In section~\ref{section4}, we provide a detailed discussion on the results and we summarize our findings in section~\ref{section5}.

\section{Event Generation and Target observables}
\label{section2}
In this section, we begin with a brief introduction on event generators. Then, we proceed to define impact parameter and transverse spherocity.

\subsection{A Multi-Phase Transport (AMPT) model}
\label{formalism}
AMPT Model contains four main components namely, initialization, followed by parton transport, hadronization mechanism and hadron transport~\cite{AMPT2}.  The initialization of the model is similar to HIJING model~\cite{ampthijing}, where the produced partons calculated in pp collisions are converted into heavy-ion collisions. They are incorporated via nuclear overlap and shadowing function using inbuilt Glauber model. The initial low-momentum partons are produced from parametrized colored string fragmentation mechanisms and they are separated from high momentum partons by a momentum cut-off. The produced partons are then initiated into parton transport part, ZPC~\cite{amptzpc}. In the String Melting version of AMPT, at the start of the ZPC melting of the colored strings into low momentum partons takes place, which is calculated using Lund FRITIOF model. Then the partons undergo multiple scatterings which take place when any two partons are within a distance of minimum approach and the transported partons are finally hadronized using spatial coalescence mechanism~\cite{Lin:2001zk,He:2017tla}. The produced hadrons further undergo final evolution in ART mechanism~\cite{amptart1, amptart2} via hadron interactions. The particle flow and spectra at the mid-$p_{\rm T}$ regions are well explained by quark coalescence mechanism for hadronization~\cite{ampthadron1,ampthadron2,ampthadron3} which is embedded in string melting mode in AMPT. Thus, we have used AMPT string melting mode (AMPT version 2.26t7) for all of our calculations. The AMPT settings in the current work are the same as reported in Refs.~\cite{Tripathy:2018bib,Mallick:2020yih}. For the input of impact parameter values for different centrality classes in Pb-Pb collisions at $\sqrt{s_{\rm NN}}$ = 2.76 and 5.02 TeV, we have used Ref.~\cite{Loizides:2017ack}. 

\subsection{PYTHIA8 (Angantyr)}
PYTHIA8~\cite{Sjostrand:2014zea}, which was initially developed for small collision systems such as $e^{+}e^{-}$, pp and $p\bar{p}$ collisions, now includes Angantyr model for the predictions for heavy-ion collisions. The main idea of Angantyr model in PYTHIA8 is to extrapolate dynamics from pp collisions to heavy-ion collisions, retaining as much as possible from pp collisions~\cite{Bierlich:2018xfw}. In order to make predictions for heavy-ion collisions, different parts of a standard PYTHIA8 simulation was modified and it was tuned with the results from $e^{+}e^{-}$, pp and $e$p collisions. So far, the model does not use any heavy-ion data to tune it. Thus, the current model retains the production mechanisms from small collision systems. However, it is successful in reproducing several features of pA and AA collisions~\cite{Bierlich:2018xfw}. In this work, we have used the predictions from PYTHIA8 (Angantyr) model to show the model dependence of the ML technique in obtaining the transverse spherocity distributions.

\subsection{Impact Parameter}
The interpretation of several results measured in heavy-ion collisions largely depends on the overlap region of two colliding nuclei in a given impact parameter ($b$). Obtaining the impact parameter values from experiments are still challenging as its value ranges in few femtometers ($fm$). However, theoretical techniques, using the so-called Glauber formalism~\cite{Glauber:1970,Miller:2007ri,Loizides:2016,Wong:1994book} have been developed to allow estimation of impact parameter and number of participants from experimental data, which consider multiple scattering of nucleons in nuclear targets. AMPT and PYTHIA8 (Angantyr) model internally depend on Glauber picture to model the early stage of heavy-ion collisions with a proper computation of the number of inelastic sub-collisions for a particular centrality class~\cite{dEnterria:2020dwq}. Here, we briefly describe how the total inelastic cross-section, number of binary collisions and number of participants are related to the impact parameter.

For a collision of two heavy-nuclei, $A$ and $B$ at relativistic speeds with impact parameter ${\bf b}$, the inelastic cross-section can be defined as


\begin{eqnarray}
\sigma^{\rm inel}_{AB}({\bf b})&=&\int d{\bf b}\left[1-\left(1-T_{AB}({\bf b})\sigma_{NN}^{\rm inel}\right)^{AB}\right]\\
&\simeq&\int d{\bf b}\left[1-\exp\left[-ABT_{AB}({\bf b})\sigma_{NN}^{\rm inel}\right]\right]\, ,
\end{eqnarray}
where, $T_{AB}({\bf b})$ is known as the nuclear overlap function and $\sigma_{NN}^{\rm inel}$ is the nucleon-nucleon inelastic cross-section. 
For such nucleus-nucleus collisions, the total number of binary collisions is
\begin{eqnarray}
N_{\rm coll}^{AB}({\bf b})&=&\sum_{n=0}^A nP(n,{\bf b})=ABT_{AB}({\bf b})\sigma_{NN}^{\rm inel}\, ,
\end{eqnarray}
and the number of participants  (or wounded nucleons) of nucleus $A$ for a given impact parameter, {\bf b} is given by
\begin{eqnarray}
N_{\rm part}^A({\bf b})=B\int {{T}_B \left({\bf{s}-\bf{b} } \right)\left\{ {1 - \left[ {1 - T_A \left( {\bf{ s} } \right)\sigma^\mathrm{NN}_\mathrm{inel} } \right]^A }\right\}d^2 s} 
\end{eqnarray}
The number of participants in nucleus $A$ is proportional to the nuclear profile function at transverse positions ${\bf s}$, $T_{AB}({\bf s})$, weighted by the sum over the probability for a nucleon-nucleon collision at transverse position (${\bf {s-b}}$) in nucleus $B$. Thus at a given ${\bf b}$, the number of participants is given by
\begin{equation}
N_{\rm part}({\bf b})=N_{\rm part}^A({\bf b})+N_{\rm part}^B({\bf b})\, .
\end{equation}
Theoretical calculations in heavy-ion physics use $\bf b$ as an input to compare theoretical results to the experimental ones. $N_{\rm part}({\bf b})$ or $N_{\rm coll}({\bf b})$ are calculated using Glauber model at a given $\bf b$, which are subsequently related to multiplicities~\cite{Kolb:2001}. In this article, we use machine learning technique to predict the impact parameter distribution using the observables measured after the collision.

\subsection{Transverse Spherocity}
Transverse spherocity is defined for a unit vector $\hat{n}$ that minimizes the following ratio in the transverse plane:
\begin{eqnarray}
S_{0} = \frac{\pi^{2}}{4} \bigg(\frac{\Sigma_{i}~|\vec p_{T_{i}}\times\hat{n}|}{\Sigma_{i}~p_{T_{i}}}\bigg)^{2}.
\label{eq1}
\end{eqnarray}

By definition, transverse spherocity is infrared and collinear safe~\cite{Salam:2009jx} and the extreme limits of transverse spherocity are related to specific configurations of events in the transverse plane. The value of transverse spherocity ranges from 0 to 1. Transverse spherocity becoming 0 means, the events have back-to-back structure and called as jetty events while 1 would mean the events are isotropic in nature. The isotropic events are the results of soft processes while the jetty events are usually the hard events. The spherocity distributions are obtained for the events with at least 5 charged particles in the pseudo-rapidity range of $|\eta|<0.8$ with $p_{\rm{T}}>$~0.15~GeV/$c$ to recreate the similar conditions as in ALICE at the LHC. In recent years, there have been several applications of transverse spherocity at the LHC energies, which can be found in Refs.~\cite{Cuautle:2014yda,Cuautle:2015kra,Salam:2009jx,Bencedi:2018ctm,Banfi:2010xy,Khuntia:2018qox,Tripathy:2019blo,Tripathy:2020jue,Tripathy:2020lla}. 

\begin{figure}[ht!]
\includegraphics[scale=0.30]{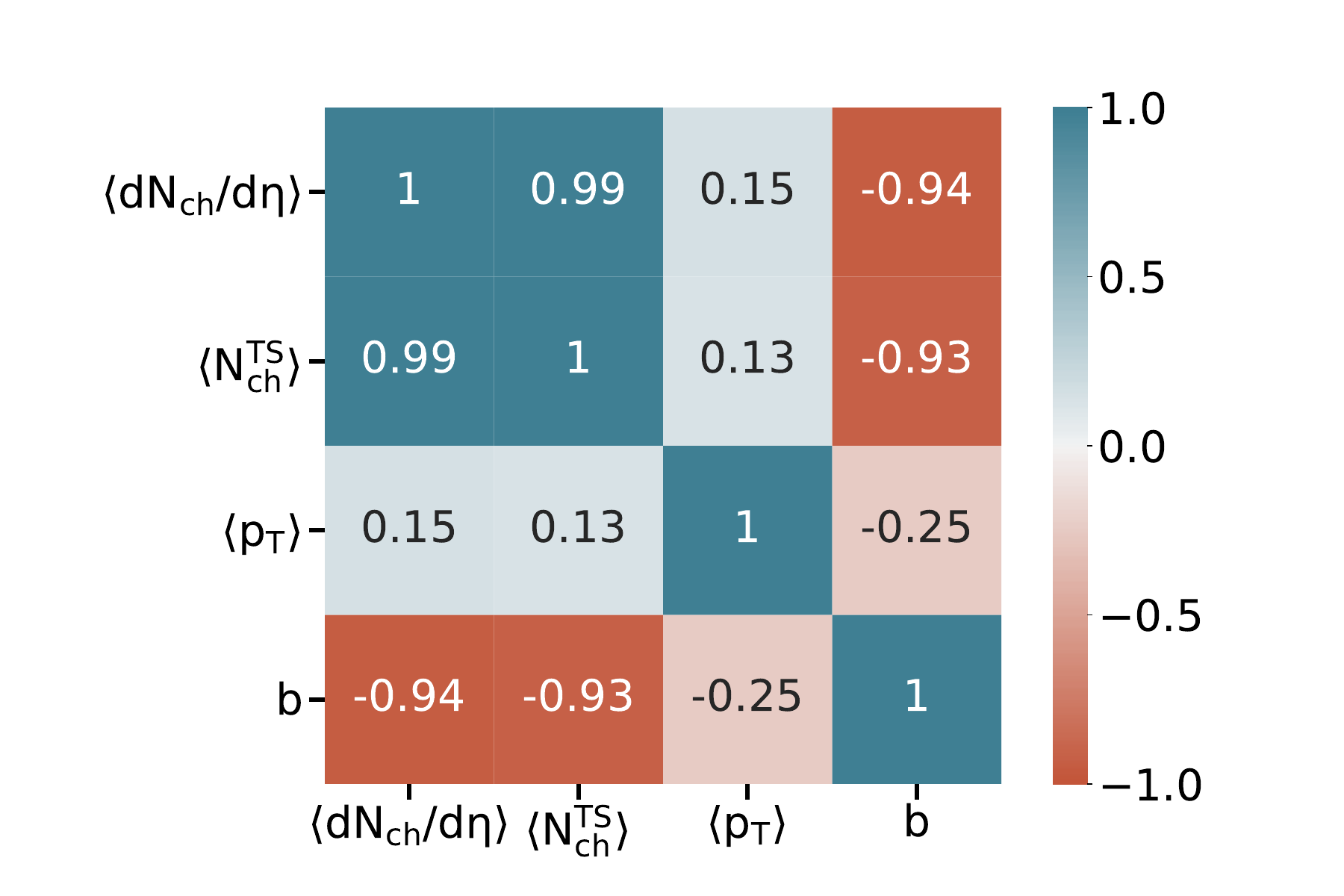}
\includegraphics[scale=0.30]{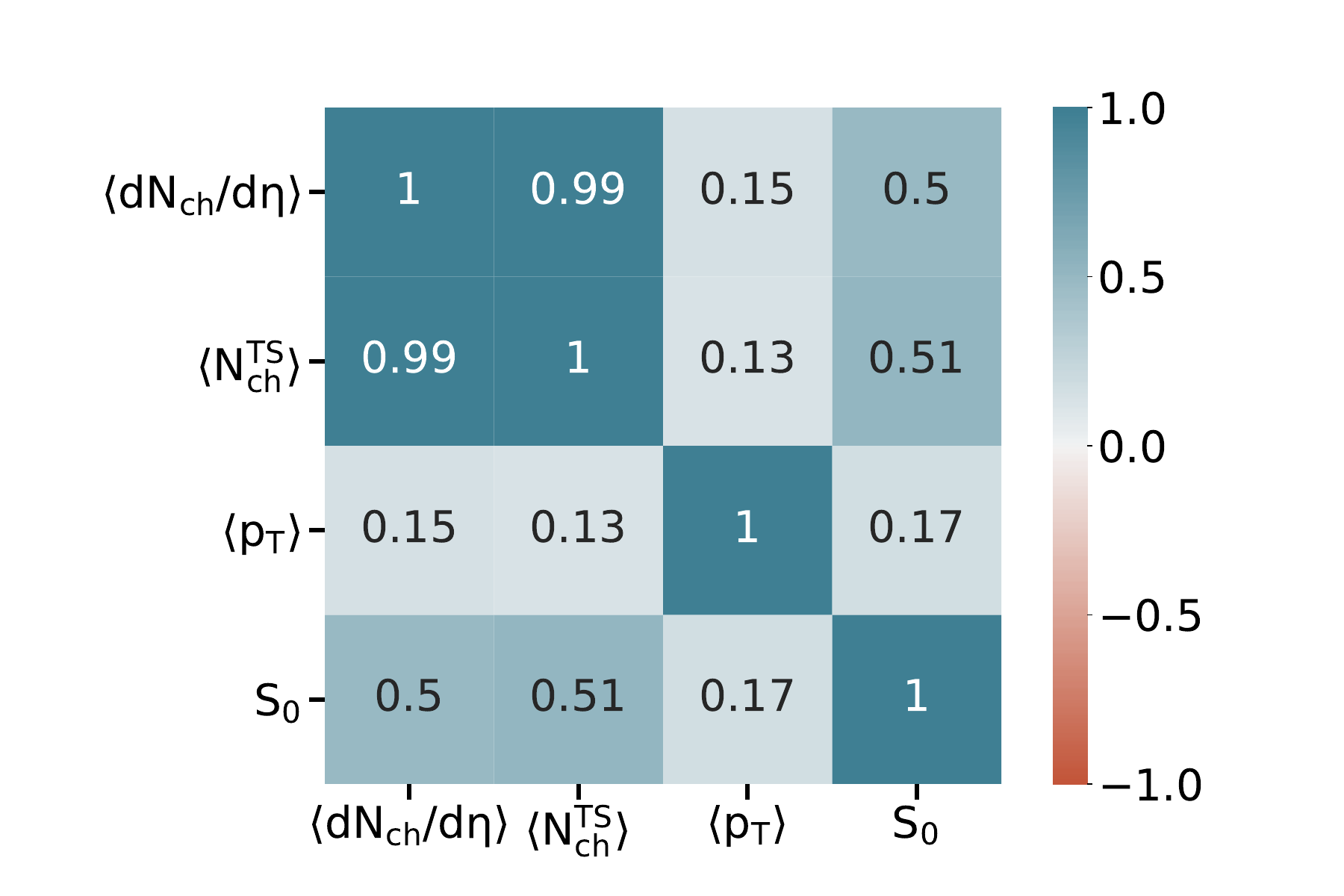}
\caption[]{(Color Online) Correlation matrix of the input variables and target observables in Pb-Pb collisions at $\sqrt{s_{\rm{NN}}} = 5.02$~TeV in AMPT model. The numbers show the correlation coefficients. The top panel shows the correlation matrix for impact parameter while the bottom panel shows the correlation matrix for transverse spherocity.}
\label{CorMatrix}
\end{figure}

\section{Machine learning based regression}
\label{section3}

\begin{figure*}[ht!]
\includegraphics[scale=0.4]{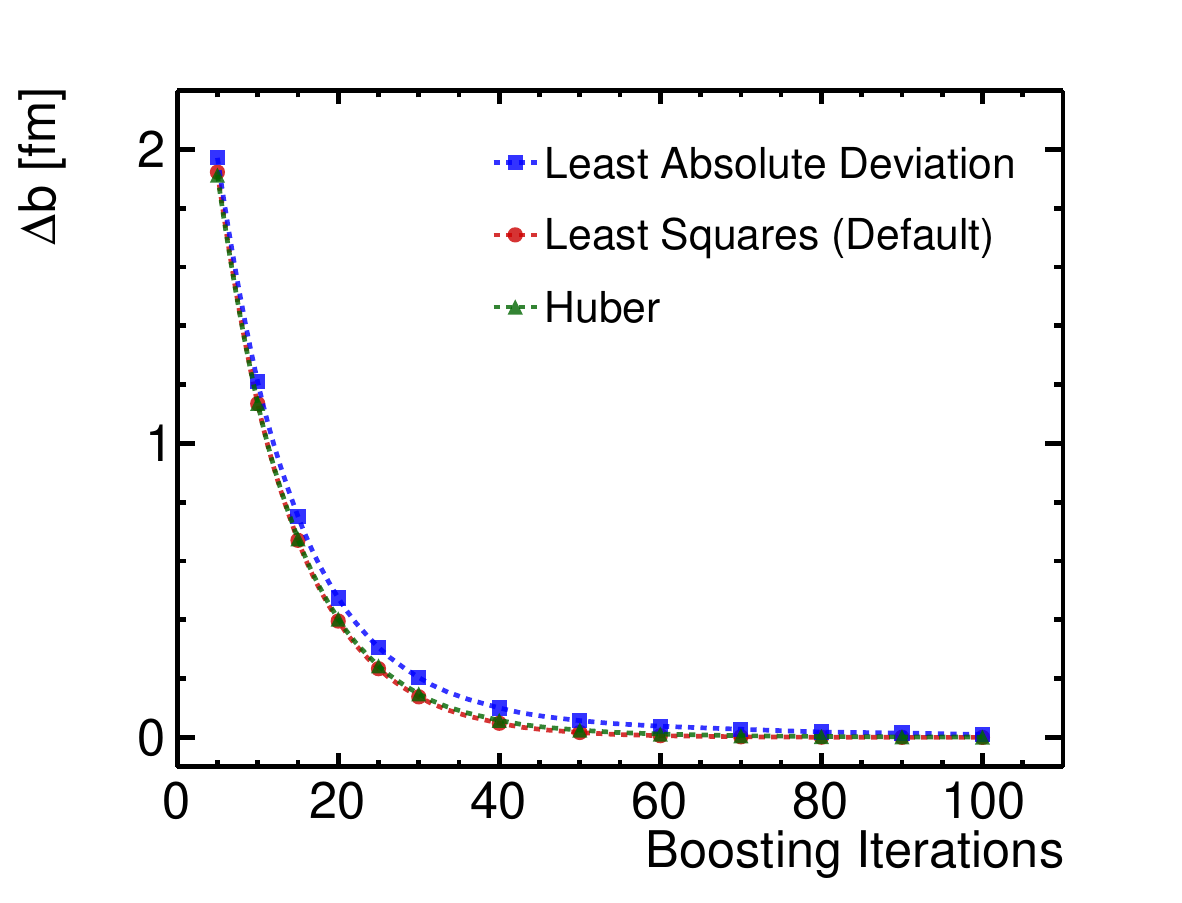}
\includegraphics[scale=0.4]{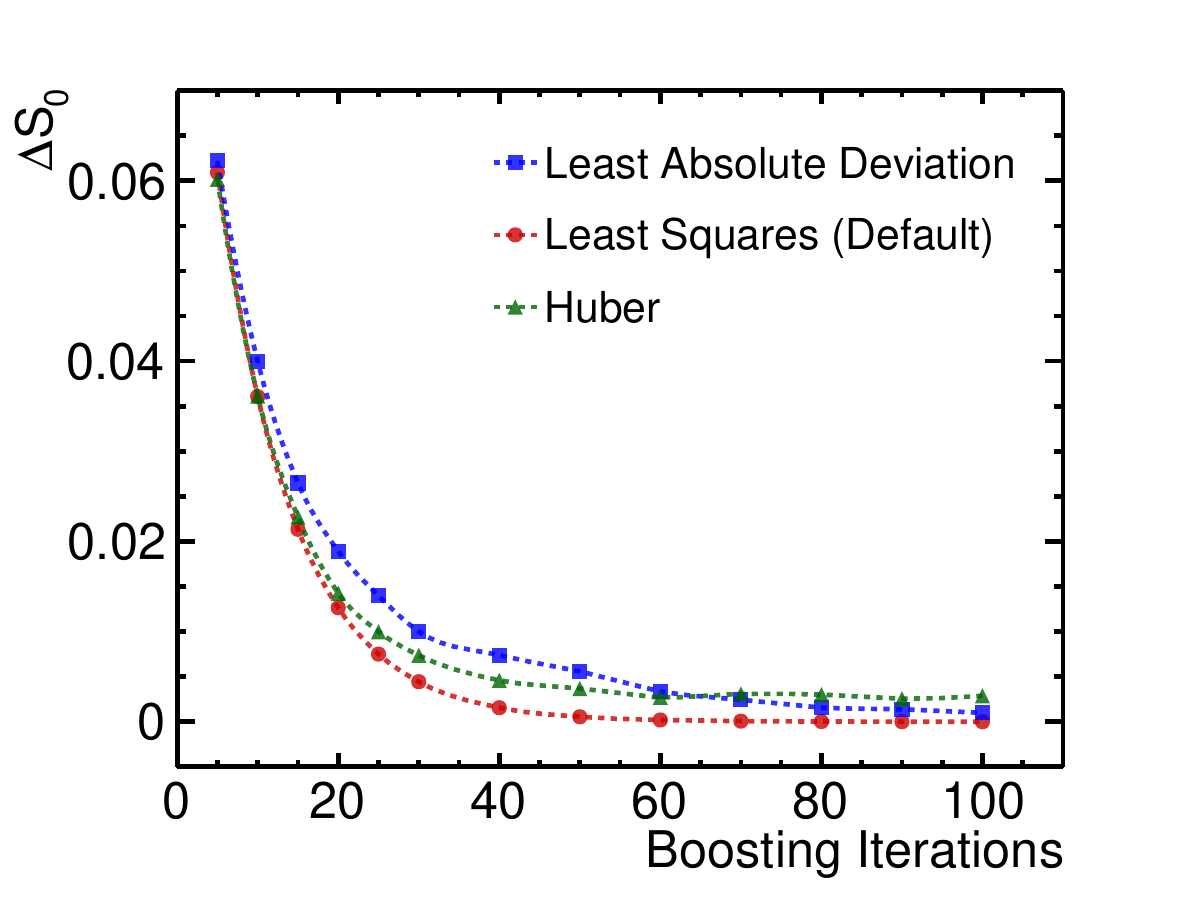}
\caption[]{(Color Online) Performance of different loss functions in GBDT-ML model in the training data set with 60K events each. The x-axis denotes boosting iterations (number of trees) and the y-axis denotes  the corresponding mean absolute error.}
\label{mae-loss}
\end{figure*}

\begin{figure*}[ht!]
\includegraphics[scale=0.4]{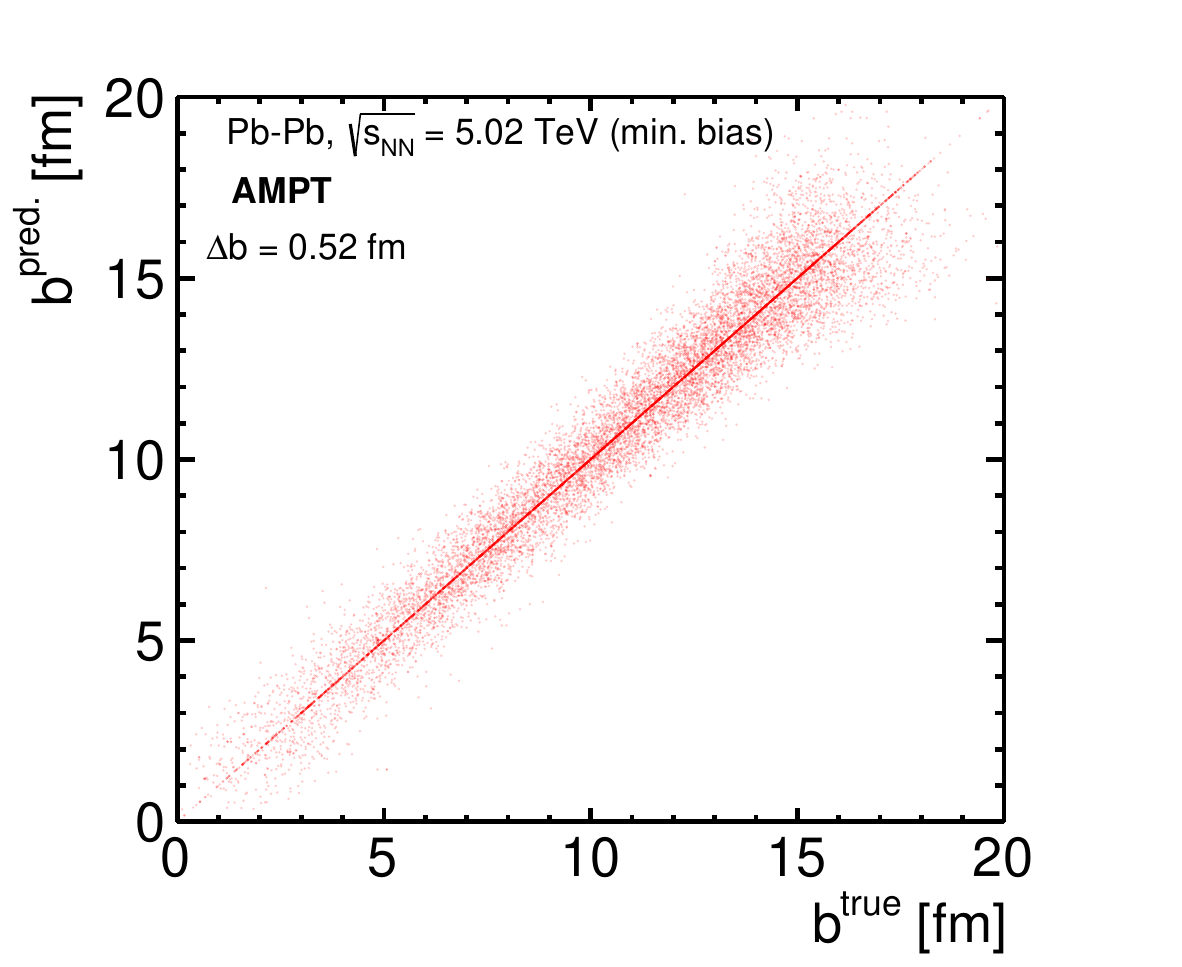}
\includegraphics[scale=0.4]{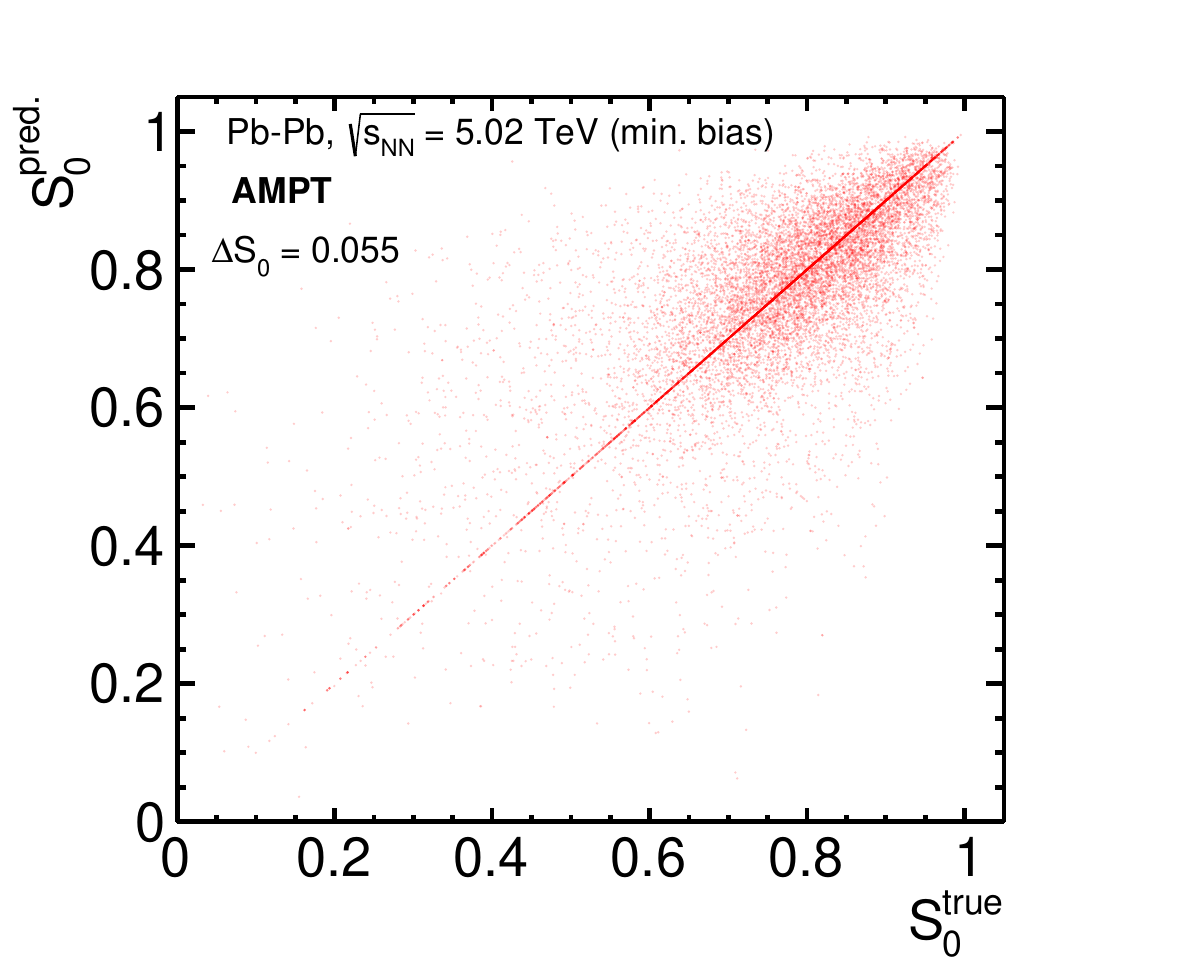}
\caption[]{(Color Online) Predicted values of impact parameter (left) and transverse spherocity (right) in the GBDT-ML model versus their true values in the testing data with 16K events of Pb-Pb collisions at $\sqrt{s_{\rm{NN}}} = 5.02$~TeV (min. bias) in AMPT model.}
\label{PredvsTrue}
\end{figure*}

Machine learning (ML) techniques could be applied to solve numerous real-life problems. Firstly, the ML model is built by training the model with a training data set. The performance is tested with a new independent set of data and further tuning of the model parameters are made if necessary. Once the predictions or the estimations are sufficiently satisfying, the model is saved and is ready to be applied to actual data to solve the problem. Machine learning addresses mainly classification, regression and clustering kind of problems. The problem, that we have in hand, is of supervised regression kind, {\em i.e.} for each set of the input variables, we have a finite numerical value as the target variable. Each set of the data refers to one event of the heavy-ion collisions. We have used charged-particle multiplicity ($\langle dN_{\rm ch}/d\eta \rangle$),  charged-particle multiplicity in the transverse region ( $\langle N_{\rm ch}^{\rm TS}\rangle$), mean transverse momentum ($\langle p_{\rm T} \rangle$) as the input variables and the target variables as the impact parameter ($b$) and transverse spherocity ($S_{0}$). For our problem, the gradient boosting decision trees (GBDT) algorithm has been chosen. Figure \ref{CorMatrix} represents the correlation matrix for the input variables and the target variables for Pb-Pb collisions at $\sqrt{s_{\rm NN}} = 5.02$ TeV minimum bias events. The numbers in the boxes represent the correlation coefficient which ranges from -1 to 1 and give the correlation strength between the intersecting variables in the matrix. The correlation coefficient ($\rho$) for two variables $x$ and $y$ is given by,
\begin{eqnarray}
\rho = \frac{{\rm cov}(x,y)}{\sigma_{x} \sigma_{y}}
\label{rho}
\end{eqnarray}
where ${\rm cov}(x,y)$ is the covariance and $\sigma_x$ and $\sigma_y$ are the standard deviations of $x$ and $y$ respectively. 

\subsection{Decision Trees Regression}
Machine learning comprises of several statistical predictive models. The algorithm learns from the data and builds the model, which then makes predictions or decisions based on its learning experience. Out of many such algorithms, decision trees are the most popular machine learning algorithms known for its simplicity yet intelligent and powerful predictions. Decision trees regression~\cite{MLbook} makes predictions for a target variable having continuous finite values (such as real numbers). In this study, these are the impact parameter ($b$) and transverse spherocity ($S_{0}$). Decision trees are built in a top-down approach. Trees can be understood as continuous piece-wise structures that take decisions based on certain rules giving rise to binary splitting of the nodes. The tree begins from the root node and then keeps on splitting recursively into further nodes. The splitting process continues till the preset maximum depth of the tree is reached. Each such split points are termed as internal nodes and the criteria of splitting is different for the type of the problem {\em i.e.} classification or regression. Splitting is often governed by minimizing the node impurity. Impurity criteria is a mathematical measure of selecting the best features for splitting and growing the tree. In decision tree regression, there are two common impurity measures {\em i.e.} least-squares and least-absolute-deviation. In least-squares, splits are chosen to minimize the sum of squared error between the observation and the mean in each node. In least-absolute-deviation, splits are chosen to minimize the mean absolute deviation from the median in each node. Mean-absolute-deviation is more robust to outliers as compared to mean-squared-error, however it fits slower. 

\begin{figure}[ht!]
\includegraphics[scale=0.35]{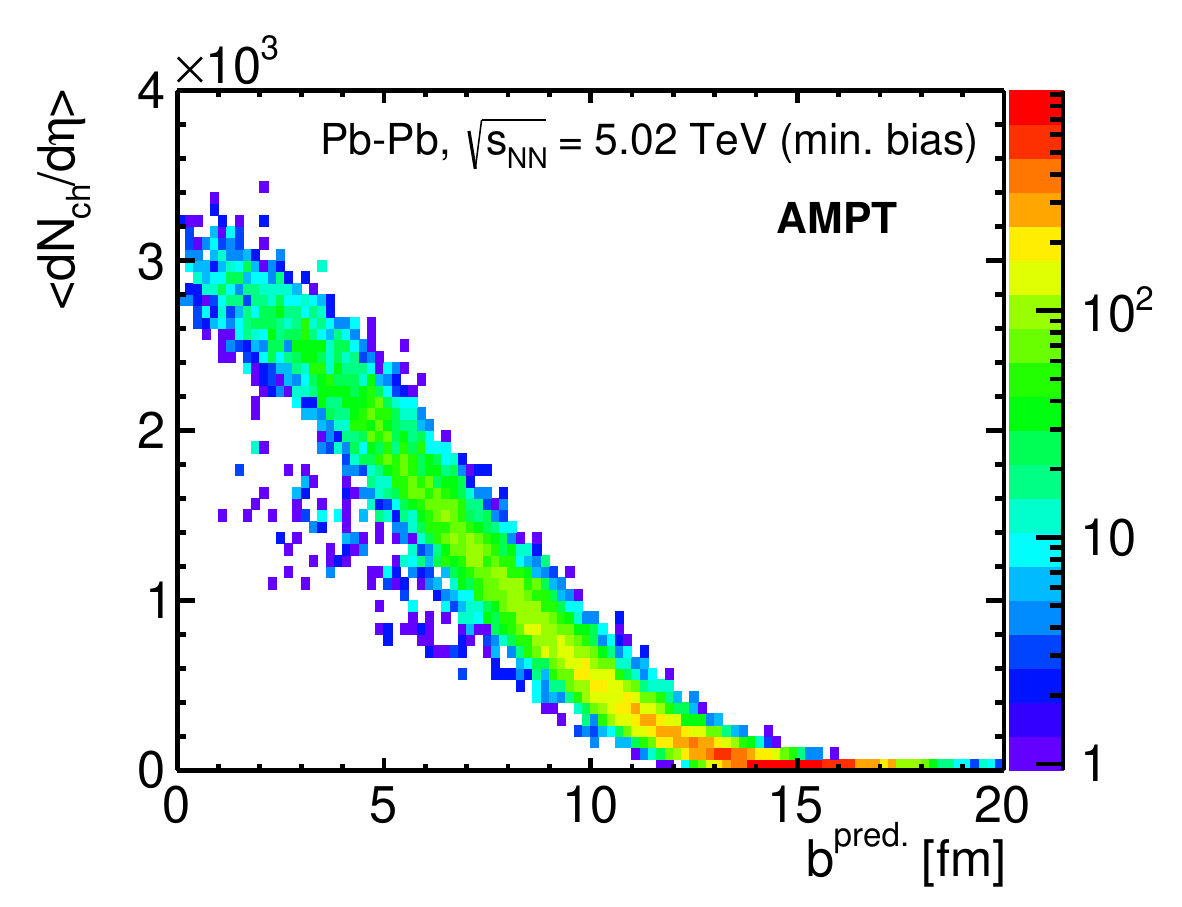}
\includegraphics[scale=0.35]{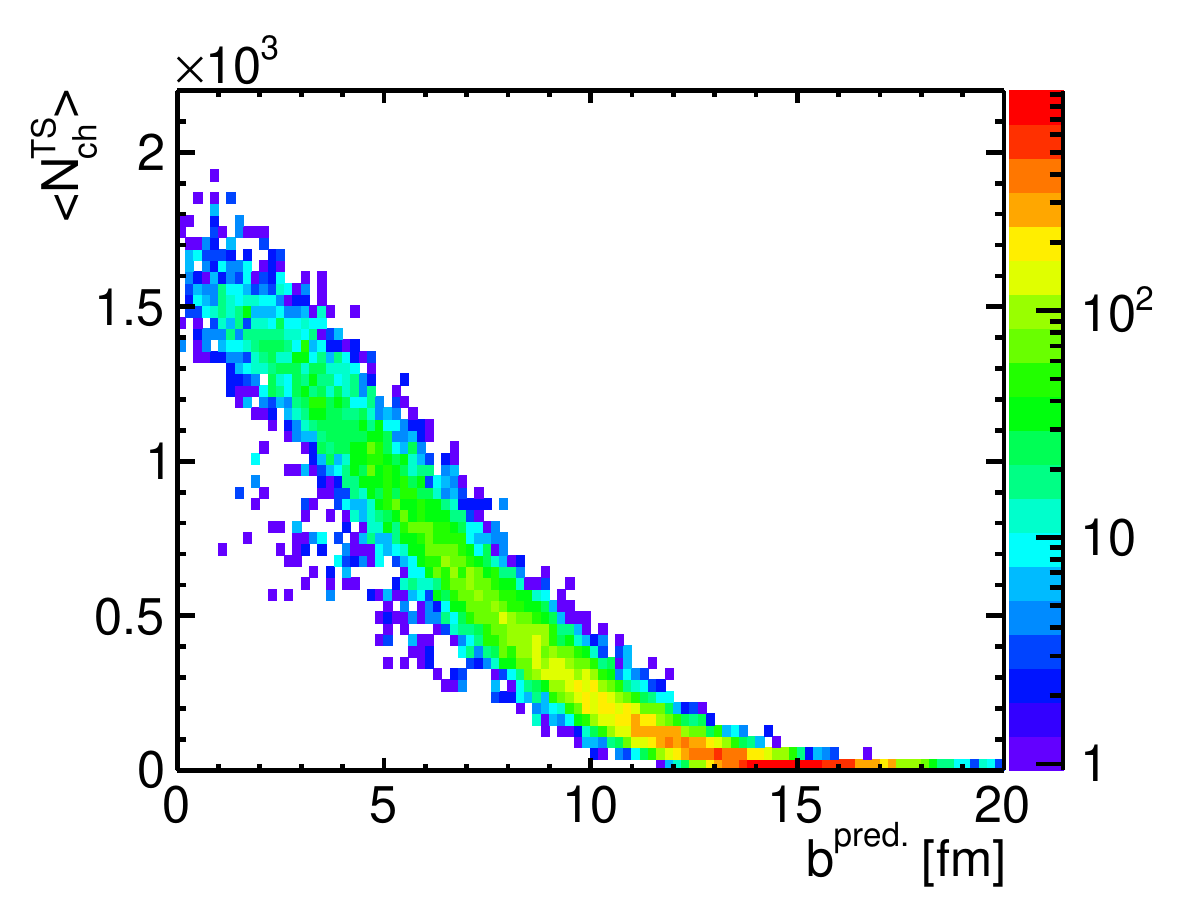}
\includegraphics[scale=0.35]{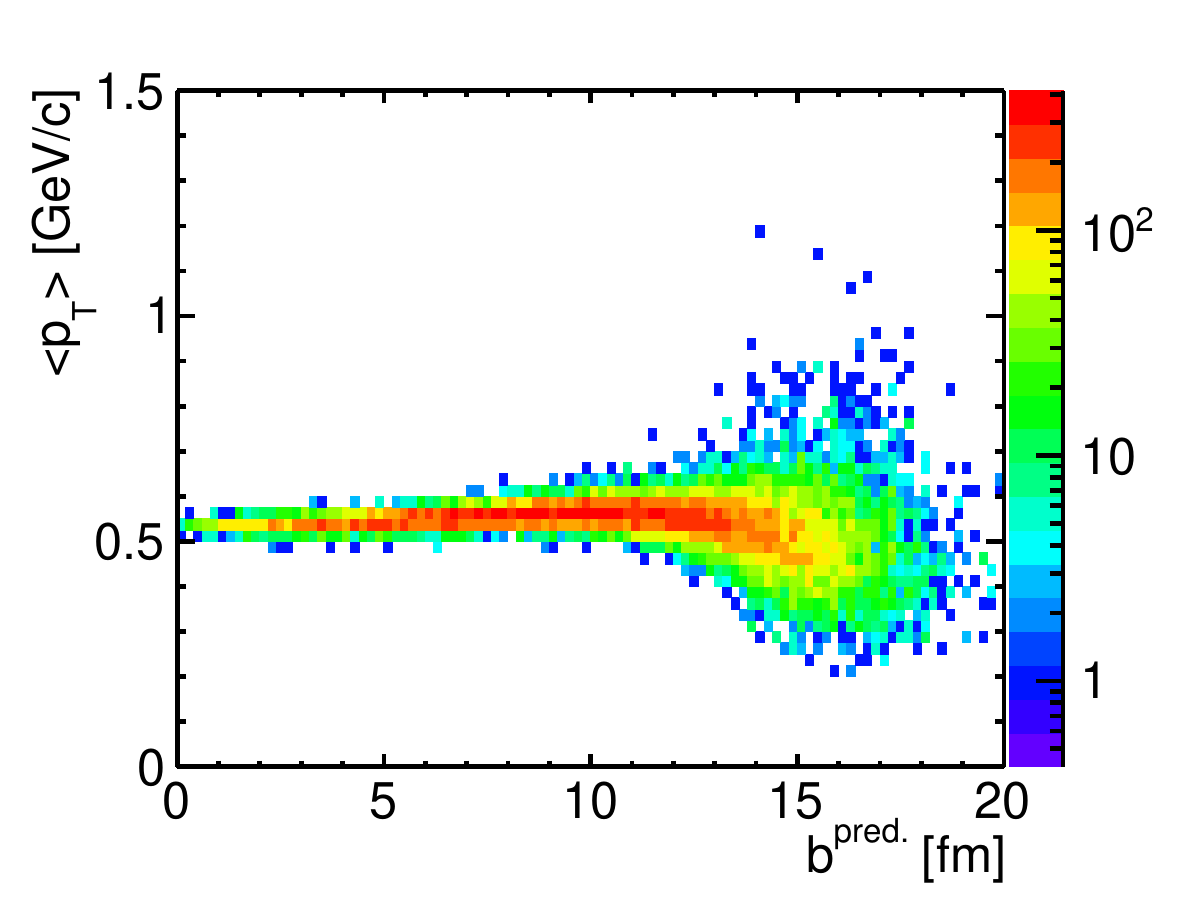}
\caption[]{(Color Online) Correlation plots between each input variable and the predicted value of impact parameter in Pb-Pb collisions at $\sqrt{s_{\rm{NN}}} = 5.02$~TeV in AMPT model.}
\label{CorrPlots-b}
\end{figure}

\subsection{Gradient Boosting Decision Trees (GBDT)}
Gradient boosting decision trees (GBDT) \cite{friedman} uses decision trees of fixed size as the base estimators. These base estimators are called as the weak learners in the context of gradient boosting. Gradient boosting is an iterative process and it builds an additive model in a forward stage-wise fashion where addition of a new weak learner compensates the shortcomings of the existing weak learners. These shortcomings are identified as the gradients [Eq. \ref{eq6}].
In any regression problem, we have a set of target variables $y$ and a set of input (observed) variables ${\bf x} = \{x_1,....,x_n\}$. The training sample $\{y_i, {\bf x_i} \}_1^N$ has all the known values of $(y,{\bf x})$ for $N-$events. The goal is to train the ML-model to obtain the functional value $F({\bf x})$ which satisfy $y_i = F({\bf x_i})$. In gradient boosting method, this estimation can be achieved by adding the outcomes of several weak learners $h_m$ as 
\begin{eqnarray}
y_i = F_M({\bf x_i}) = \sum\limits_{m=1}^{M} h_m({\bf x_i})
\label{eq3}
\end{eqnarray}
The parameter $M$ corresponds to the number of trees in each decision tree estimator. Now at each stage, the additive process can be written as 
\begin{eqnarray}
F_{m}({\bf x}) = F_{m-1}({\bf x})+\nu h_m({\bf x})
\label{eq4}
\end{eqnarray}
The parameter $\nu$ is called as the learning rate. There is a direct trade-off between the learning rate and number of trees (the number of weak learners), specified by the parameter $M$. Smaller values of learning rate require larger numbers of weak learners (more number of trees) to maintain a constant training error. Usually a model is built with a small value of learning rate as it performs better and achieves minimal testing error. The newly added tree $h_m({\bf x})$ is fitted in order to minimize the sum of a loss function $l(y_i,F_{m-1}({\bf x_i})+h_m({\bf x_i}))$.
We can use the Taylor's first order expansion and approximate the value of $l$ as
\begin{multline}
l(y_i, F_{m-1}({\bf x_i}) + h_m({\bf x_i})) \approx
l(y_i, F_{m-1}({\bf x_i})) \\
+ h_m({\bf x_i})
\left[ \frac{\partial l(y_i, F({\bf x_i}))}{\partial F({\bf x_i})} \right]_{F=F_{m - 1}}.
\label{eq5}
\end{multline}
For the squared loss function which is of the form, 
$l(y_i,F({\bf x_i})) = \frac{1}{2}(y_i-F({\bf x_i}))^2$,
\begin{multline}
-g_i = -\left[ \frac{\partial l(y_i, F({\bf x_i}))}{\partial F({\bf x_i})} \right]_{F=F_{m - 1}} = y_i-F({\bf x_i})
\label{eq6}
\end{multline}
Here, $g_i$ is the gradient and $(y_i-F({\bf x_i}))$ is called as the residual. We can interpret residuals as negative gradients. Now, to improve the model predictions and build more robust model, a suitable loss function is chosen, which is then minimized using the gradient descent algorithm. In GBDT algorithm, we have three types of loss functions {\em i.e.} least-squares, least-absolute-deviation and the Huber loss functions.\\
\\{\bf Least-squares:}
$$ l(y_i,F({\bf x_i})) = \frac{1}{2}(y_i-F({\bf x_i}))^2$$
{\bf Least-absolute-deviation:}
$$ l(y_i,F({\bf x_i})) = |y_i-F({\bf x_i})|$$
{\bf Huber:}
$$l(y_i,F({\bf x_i})) =
\begin{cases}
\frac{1}{2}(y_i-F({\bf x_i}))^2, ~|y_i-F({\bf x_i})| \leq \delta \\
\delta (|y_i-F({\bf x_i})| - \delta/2), ~|y_i-F({\bf x_i})| > \delta
\end{cases}
$$
Here, $\delta$ is known as the transition point that defines those residual values that are considered to be “outliers”, subject to absolute rather than squared-error loss. For residual less than or equal to $\delta$, the Huber loss function becomes the least-squares loss.

\begin{figure}[ht!]

\includegraphics[scale=0.35]{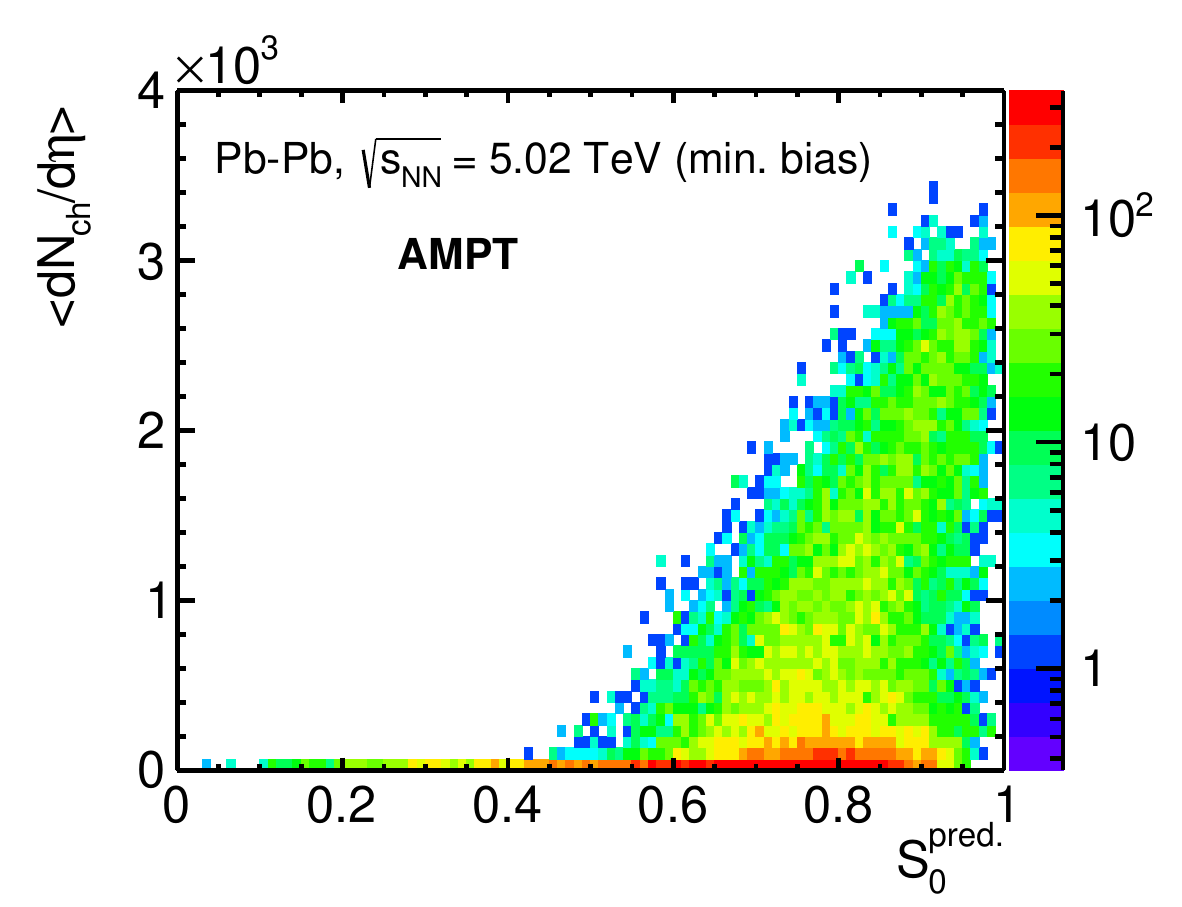}
\includegraphics[scale=0.35]{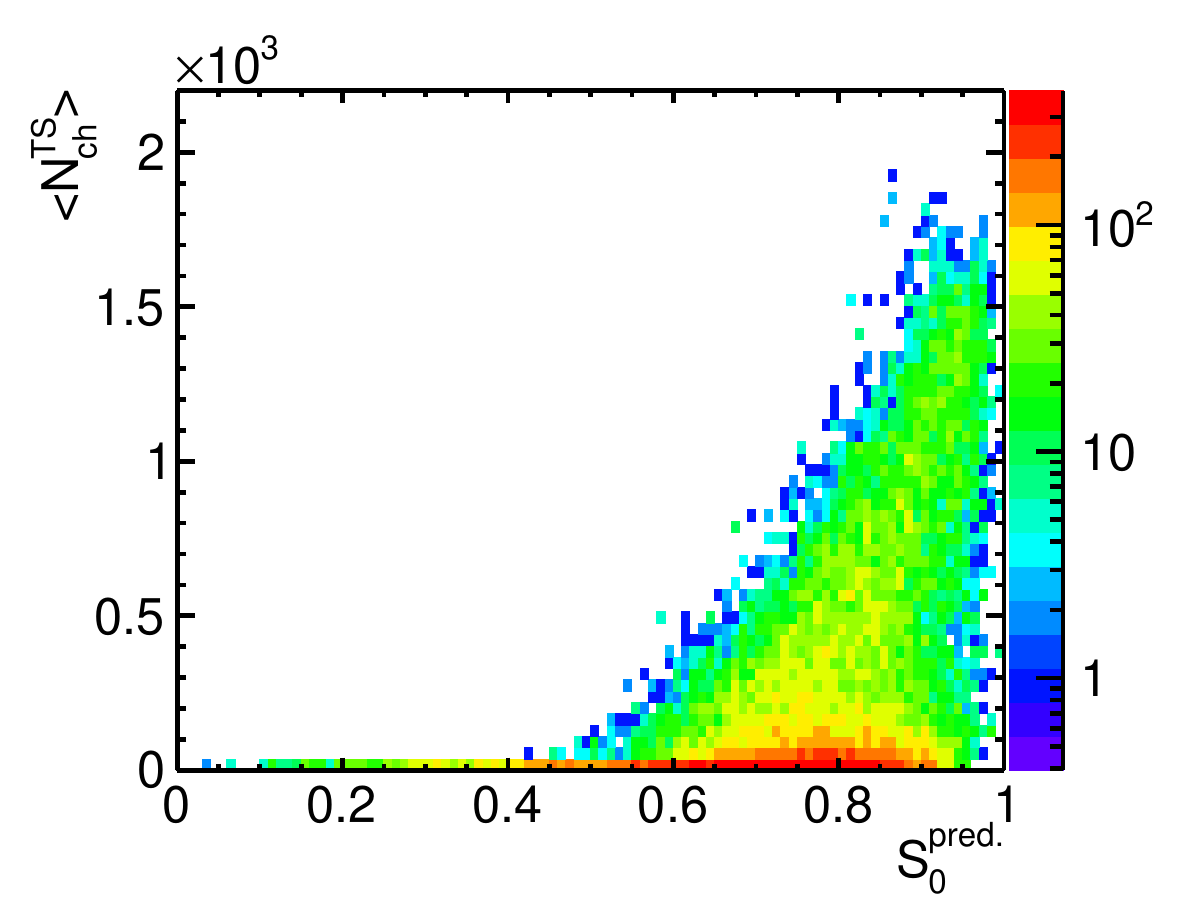}
\includegraphics[scale=0.35]{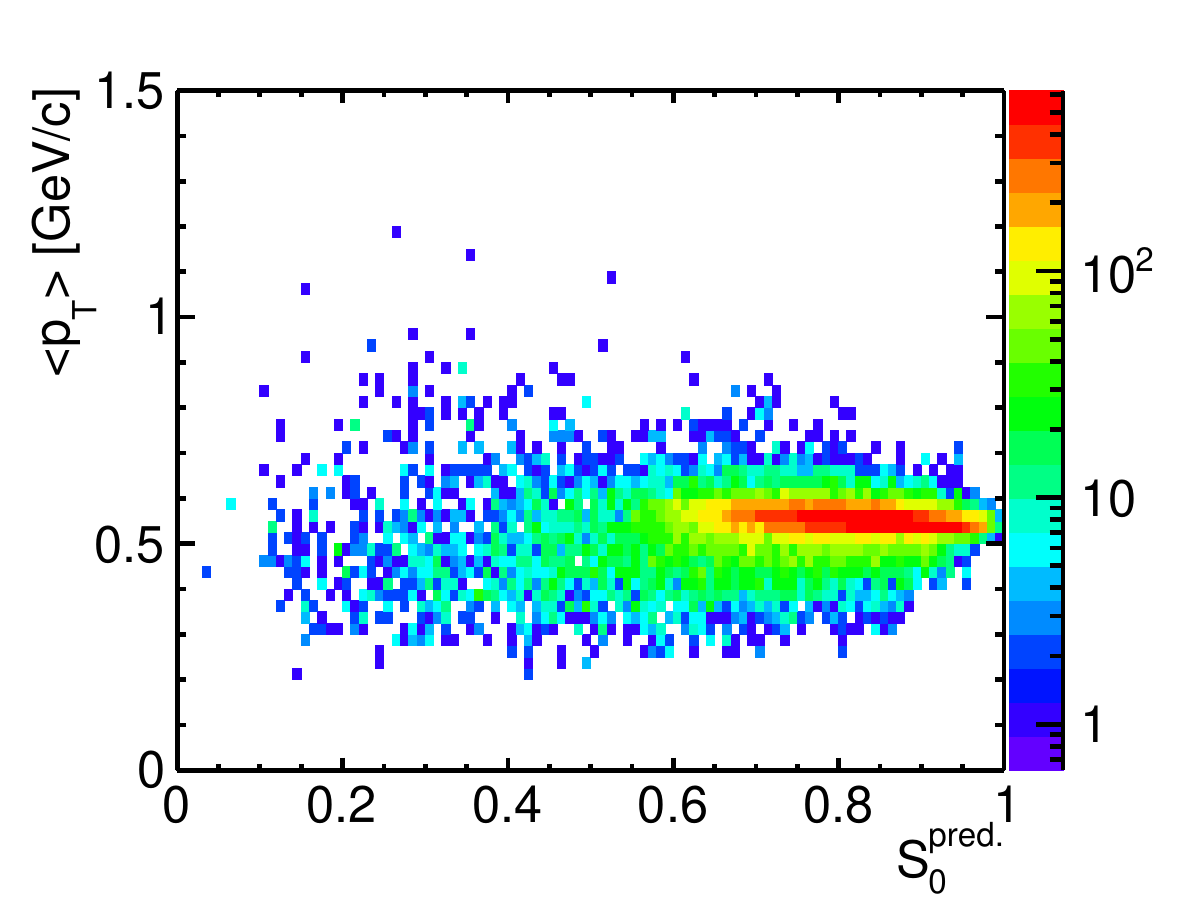}
\caption[]{(Color Online) Correlation plots between each input variable and the predicted value of transverse spherocity in Pb-Pb collisions at $\sqrt{s_{\rm{NN}}} = 5.02$~TeV in AMPT model.}
\label{CorrPlots-sphero}
\end{figure}

\subsection{Quality assurance}
In GBDT algorithm, there are essential parameters such as number of decision trees, maximum depth and learning rate, which play crucial role in the fitting process. 
The task is to obtain the best fit of the model to the training data by optimizing these parameters. These parameters require manual tuning by observing the performance of the model. For this study, we have taken the number of trees to be 100, maximum depth to be 40, learning rate is fixed at 0.1 and all other parameters are set to their default values. The accuracy of the trained model could be evaluated by calculating the mean-absolute-error of the target variables for the training data set. The mean-absolute-error for impact parameter is given by
\begin{eqnarray}
\Delta b = \frac{1}{N_{\rm events}}\sum\limits_{n=1}^{N_{\rm events}}|b_{n}^{true}-b_{n}^{pred.}|
\label{eq7}
\end{eqnarray}
Here, $b_{n}^{true}$ is the true value of the impact parameter from the simulated data and $b_{n}^{pred}$ is the predicted value from the GBDT-ML model. Mean-absolute-error for transverse spherocity ($\Delta S_{0}$) could be estimated in the similar fashion. The learning process of the ML-model is greatly influenced by the size of the training data. We can see this by evaluating the values of $\Delta b$ and $\Delta S_{0}$ for 10K events of independent testing data with the ML-model trained with different set of events. The results are mentioned in Table \ref{Tab2}. As we can see, with more number of events in the training data, the model learns better, hence the mean-absolute-error decreases with increase in training data size. This behavior is expected as the model should gather more information with more training data, and thus its prediction gets improved. As we increase training data size from 2K to 60K events, $\Delta b$ decreases from 0.71 fm to 0.52 fm and $\Delta S_{0}$ decreases from 0.079 to 0.055. However, with training size greater than 50K events, $\Delta b$ saturates to a constant value, and the decrease in $\Delta S_{0}$ is too small to make any difference. Therefore, for this study, we have taken 60K events for the training of the model for both the target variables.
\begin{table}[ht!]
\caption{Size dependence of GBDT-ML model for the training data in Pb-Pb collisions at $\sqrt{s_{\rm NN}}=5.02$ TeV. $\Delta b$ and $\Delta S_{0}$ are the mean-absolute-error on the independent testing data having 10K events for impact parameter and transverse spherocity, respectively.} 
\begin{tabular}{lllllll}
\hline
\multicolumn{1}{|l|}{Size of training data} & \multicolumn{1}{l|}{2K} & \multicolumn{1}{l|}{10K} & \multicolumn{1}{l|}{20K} &  \multicolumn{1}{l|}{40K} & \multicolumn{1}{l|}{50K} & \multicolumn{1}{l|}{60K} \\ \hline
\multicolumn{1}{|l|}{$\Delta b$ [fm]}   & \multicolumn{1}{l|}{0.71} & \multicolumn{1}{l|}{0.62}  & \multicolumn{1}{l|}{0.58}  &  \multicolumn{1}{l|}{0.53}  & \multicolumn{1}{l|}{0.52}  & \multicolumn{1}{l|}{0.52}  \\ \hline
\multicolumn{1}{|l|}{$\Delta S_{0}$}   & \multicolumn{1}{l|}{0.079} & \multicolumn{1}{l|}{0.068}  & \multicolumn{1}{l|}{0.062}  &  \multicolumn{1}{l|}{0.058}  & \multicolumn{1}{l|}{0.056}  & \multicolumn{1}{l|}{0.055}  \\ \hline
           
\end{tabular}
\label{Tab2}
\end{table}

\begin{figure}[ht!]
\includegraphics[scale=0.38]{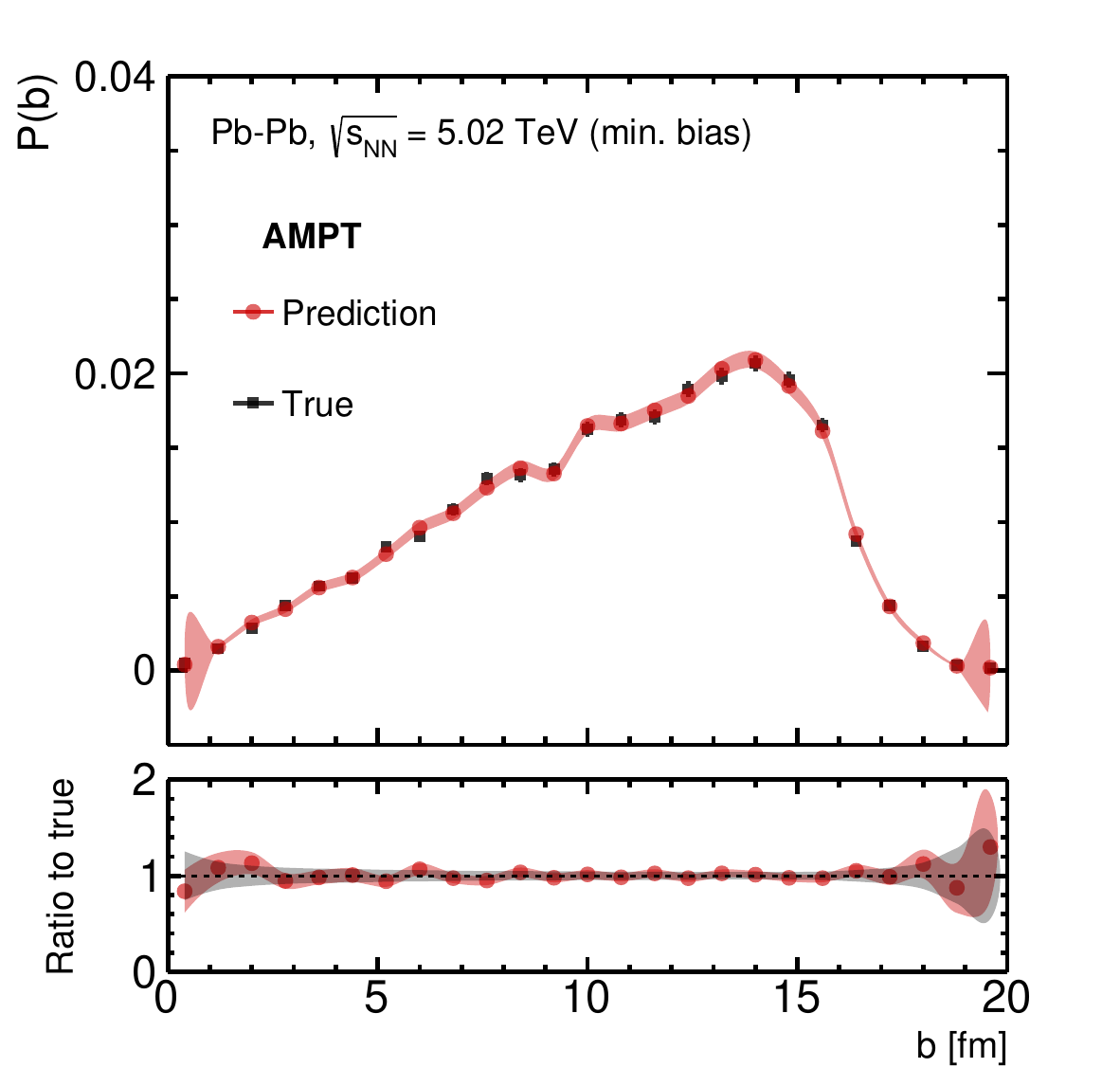}
\caption[]{(Color Online) Predictions for impact parameter distribution using gradient boosting decision trees algorithm for Pb-Pb collisions at $\sqrt{s_{\rm{NN}}} = 5.02$~TeV in AMPT model. The quadratic sum of the statistical and systematic uncertainties are shown as a red-colored band for the predicted values. The statistical uncertainties in the true values are shown as bars. In the ratio, black-colored band denotes the statistical uncertainties in the true values while the red-colored band denotes the quadratic sum of statistical and systematic uncertainties.}
\label{ImpactPar-AMPT}
\end{figure}

Loss function is another important parameter in the GBDT algorithm, which needs to be chosen carefully. We have obtained the $\Delta b$ and $\Delta S_{0}$ values against boosting iterations (number of trees) for three kinds of loss functions {\em i.e.} least-squares, least-absolute-deviation and the Huber loss functions. Figure \ref{mae-loss} shows the performance of these loss functions in the training data set with 60K events for both the target variables. The x-axis denotes boosting iterations and the y-axis denotes the corresponding mean-absolute-error of the training data. As we can see, $\Delta b$ and $\Delta S_{0}$ decrease by growing more trees in the model. The values of $\Delta b$ and $\Delta S_{0}$ fall exponentially moving from 10 to 60 number of trees and then the descent is very small. For boosting iterations greater than 80, the mean-absolute-error seems to saturate and remains fairly constant. Small values of $\Delta b$ and $\Delta S_{0}$ in the training data indicate that the model is  learning better. To be fair, we stop at 100 trees. Among these three loss functions, the least-square loss performs better and its training is faster than the mean-absolute-deviation and the Huber loss. So, we have chosen the least-square loss function as a default method for this study.

By fixing the training data sample size, optimizing the model parameters and minimizing the mean-absolute-errors {\em i.e.} $\Delta b$ and $\Delta S_{0}$ in the training data, now it is time to look into the performance of the trained model. We can predict the values of impact parameter and transverse spherocity using this ML-model. Figure \ref{PredvsTrue} shows the predicted values of the variables using ML-model versus the true values of the variables from AMPT model simulation for 16K events of minimum bias Pb-Pb collisions at $\sqrt{s_{\rm{NN}}} = 5.02$~TeV in an independent testing data set. For most accurate predictions, the points on the plot should populate a straight line inclined at an angle 45 degrees to the x-axis. Though we see a little spread in the plots, the straight lines are distinctly visible, suggesting a good agreement between the predictions from ML-model and true values from the simulation. We have also computed the testing accuracy and found that, for impact parameter, $\Delta b$ = 0.52 fm while for transverse spherocity, $\Delta S_{0}$ = 0.055 for the testing data.

For the subsequent plots, the total number of accepted events for minimum bias Pb-Pb collisions at $\sqrt{s_{\rm{NN}}} = 5.02$~TeV are 76.5K, out of which 60K events are used for the ML training (to minimize the mean-absolute-errors) and rest of the events are used for the testing purpose. The maximum deviation among ML prediction from different loss functions with respect to least-square loss function method (Default method) is used as systematic uncertainties in the predicted values. They are summed in quadrature with the statistical uncertainties and shown as red-colored band in the plots. The statistical uncertainties in the true values are shown as bars. In predicted to true ratio plots, black-colored band denotes the statistical uncertainties in the true values while the red-colored band denotes the quadratic sum of statistical and systematic uncertainties.


\begin{figure}[ht!]
\includegraphics[scale=0.38]{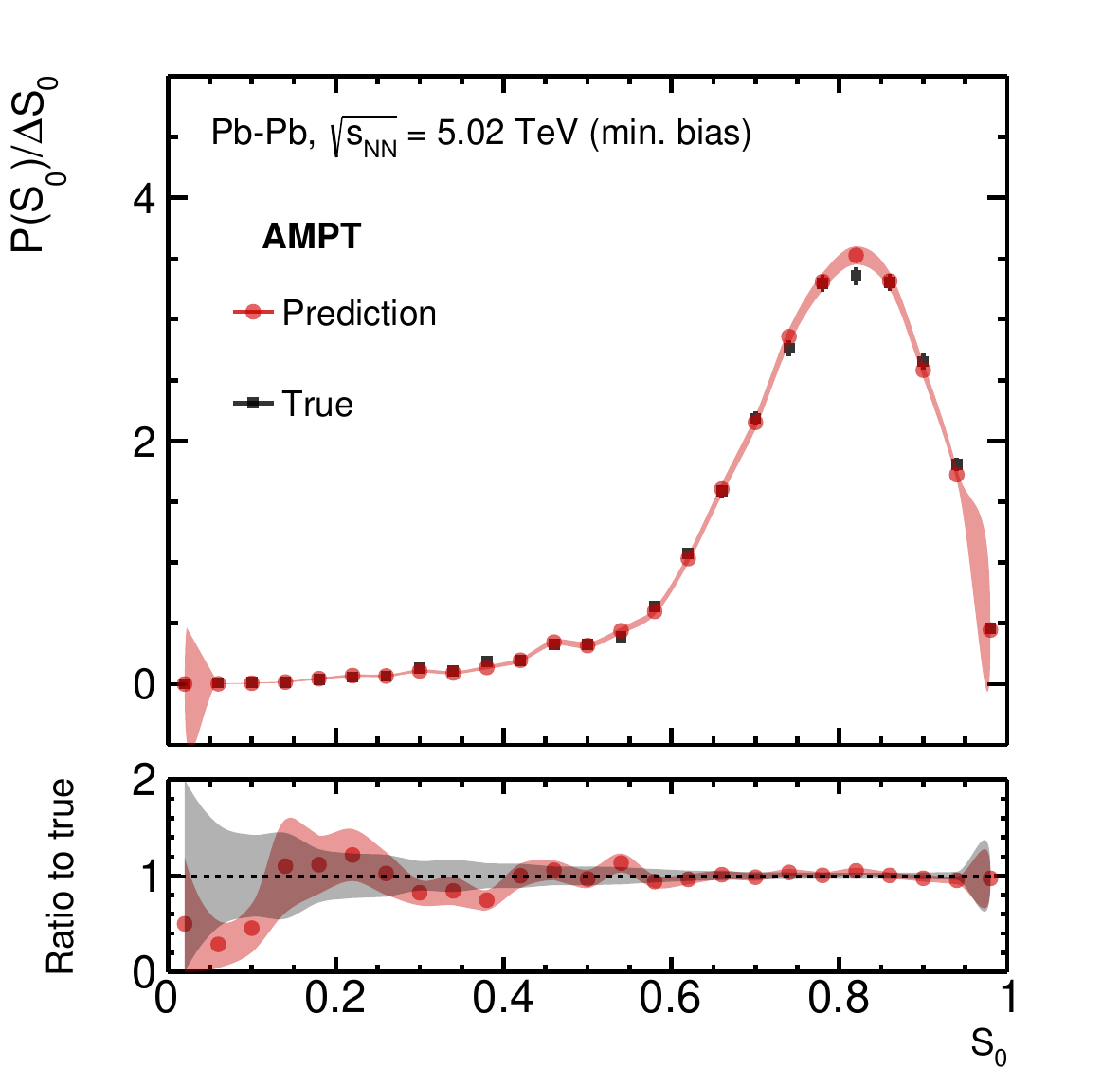}
\includegraphics[scale=0.38]{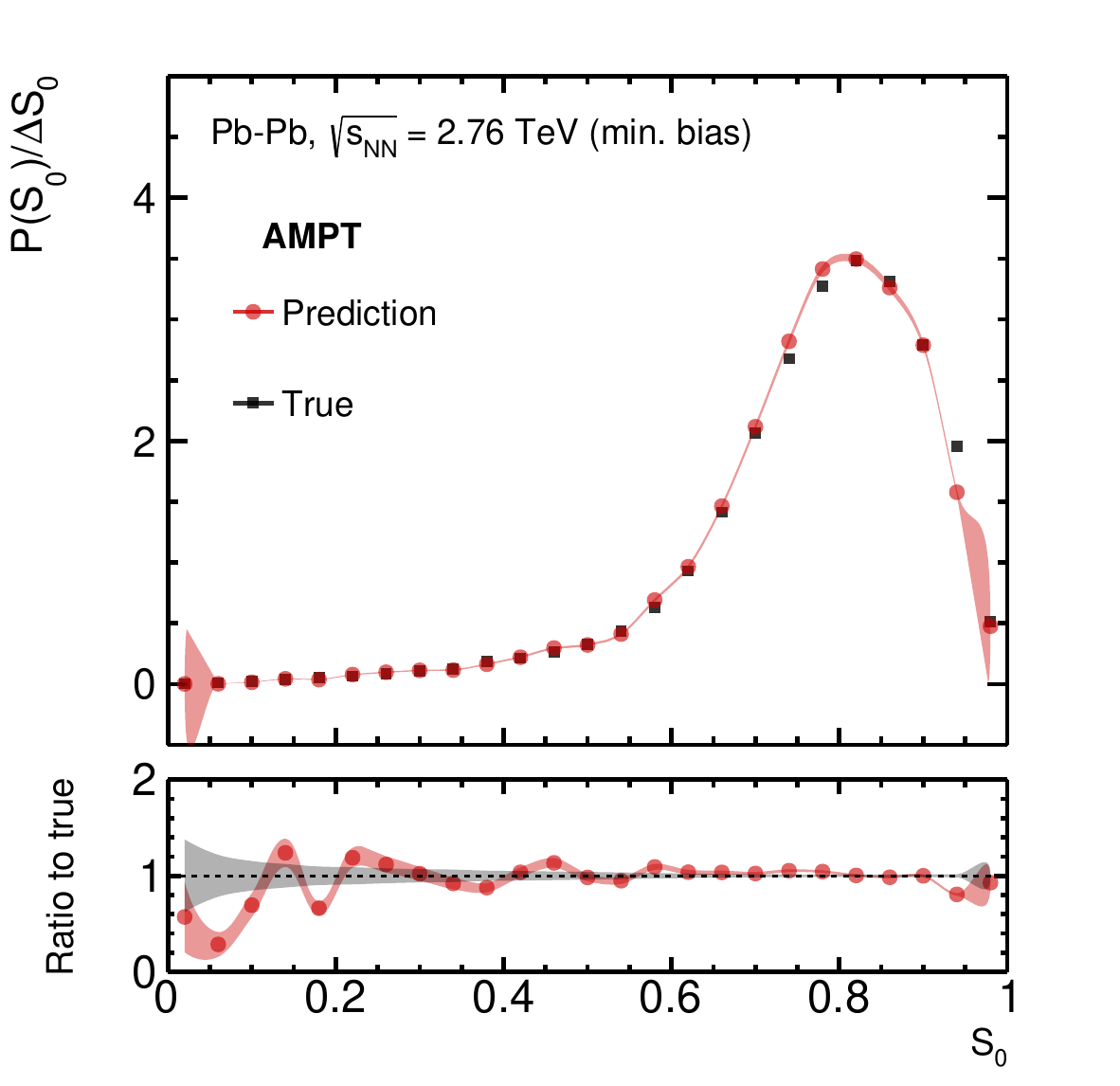}
\caption[]{(Color Online) Predictions for transverse spherocity distribution using ML and their comparison with true values in Pb-Pb collisions at $\sqrt{s_{\rm{NN}}} = 5.02$~TeV (top) and $\sqrt{s_{\rm{NN}}} = 2.76$~TeV (bottom) in AMPT model. The lower panels show the ratio of the predicted values to the true values. The quadratic sum of the statistical and systematic uncertainties are shown as a red-colored band for the predicted values. The statistical uncertainties in the true values are shown as bars. In the ratio, black-colored band denotes the statistical uncertainties in the true values while the red-colored band denotes the quadratic sum of statistical and systematic uncertainties.}
\label{Sphero-AMPT}
\end{figure}

\section{Results and Discussions}
\label{section4}

Top panel of Fig.~\ref{CorMatrix} shows the correlation matrix of the input variables and impact parameter in Pb-Pb collisions at $\sqrt{s_{\rm{NN}}} = 5.02$~TeV. The numbers show the correlation coefficients ($\rho$) which is obtained from Eq.~\ref{rho}. We see a significant anti-correlation of impact parameter and the final state charged particle multiplicities, which is evident from the values of $\rho$. Also, impact parameter was found to be (anti-)correlated with the mean transverse momentum of an event. This behavior is evident in Fig.~\ref{CorrPlots-b}, where the correlation with each input variable with impact parameter is shown. Figure~\ref{ImpactPar-AMPT} shows the predictions for impact parameter distribution using ML for Pb-Pb collisions at $\sqrt{s_{\rm{NN}}} = 5.02$~TeV in AMPT model. The lower panel shows the ratio of predicted distribution to the true distribution. One can clearly see that the proposed ML framework with $\langle dN_{\rm ch}/d\eta \rangle$,   $\langle N_{\rm ch}^{\rm TS}\rangle$ and $\langle p_{\rm T} \rangle$ as the input variables, does a nice job of predicting the impact parameter distribution in Pb-Pb collisions at $\sqrt{s_{\rm{NN}}} = 5.02$~TeV.

\begin{figure}[ht!]
	\centering
	\includegraphics[scale=0.38]{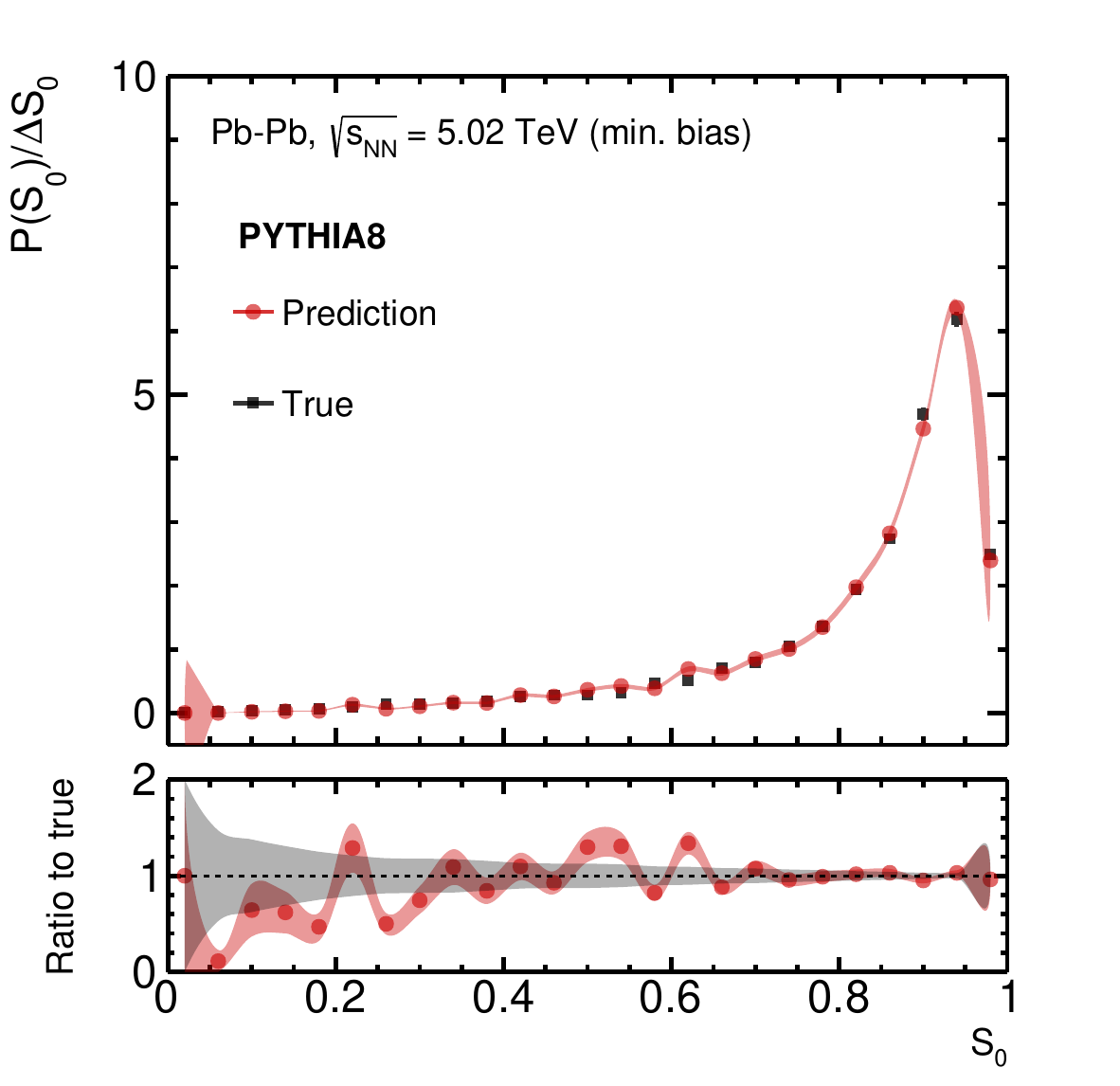}
	\caption[]{(Color Online) Predictions for transverse spherocity distribution using gradient boosting decision trees algorithm for Pb-Pb collisions at $\sqrt{s_{\rm{NN}}} = 5.02$~TeV in PYTHIA8 model. The lower panel shows the ratio of the predicted values to the true values. The quadratic sum of the statistical and systematic uncertainties are shown as a red-colored band for the predicted values. The statistical uncertainties in the true values are shown as bars. In the ratio, black-colored band denotes the statistical uncertainties in the true values while the red-colored band denotes the quadratic sum of statistical and systematic uncertainties.}
	\label{Sphero-PYTHIA}
\end{figure}

\begin{figure*}[ht!]
	\includegraphics[scale=0.38]{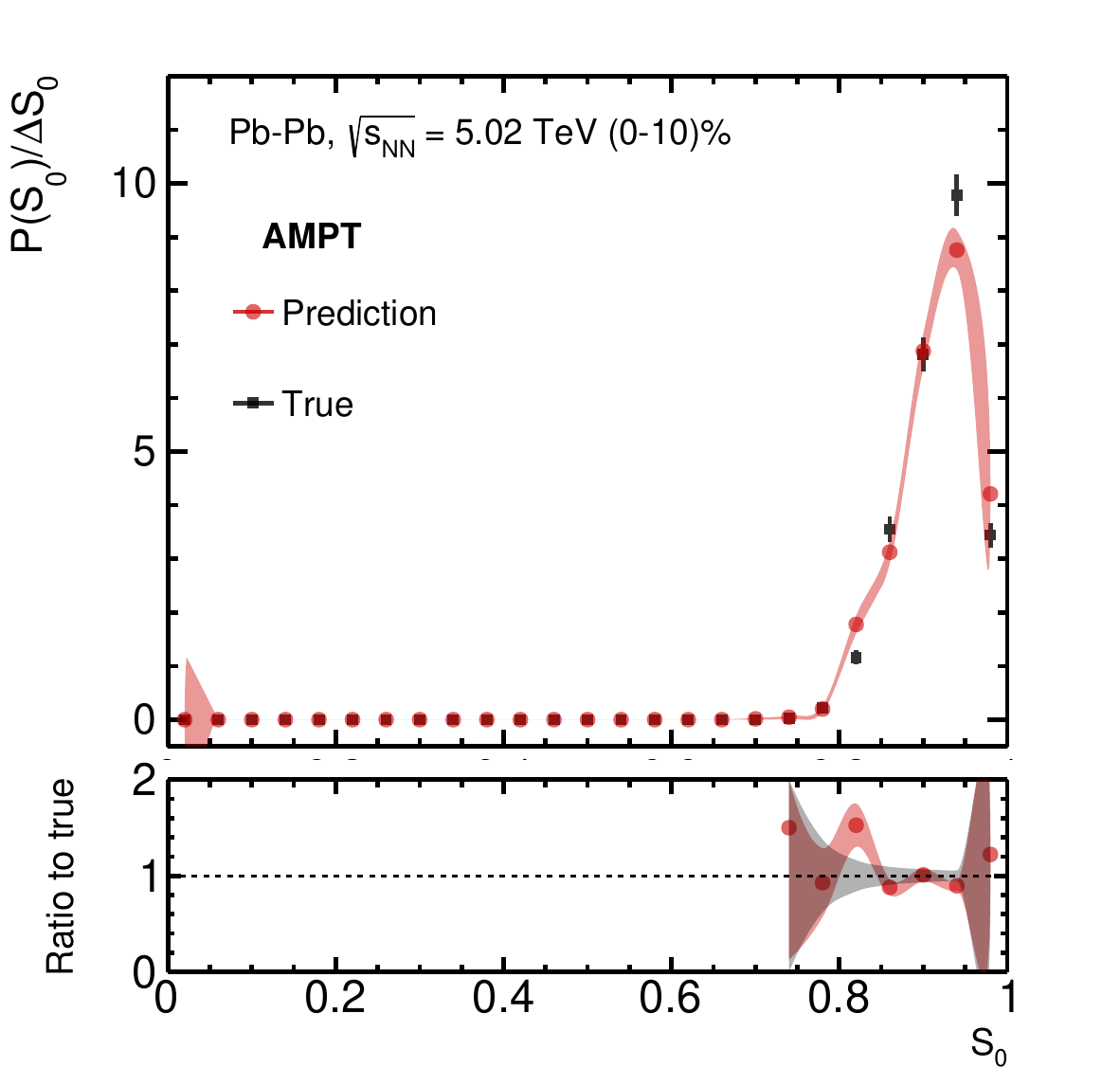}
	\includegraphics[scale=0.38]{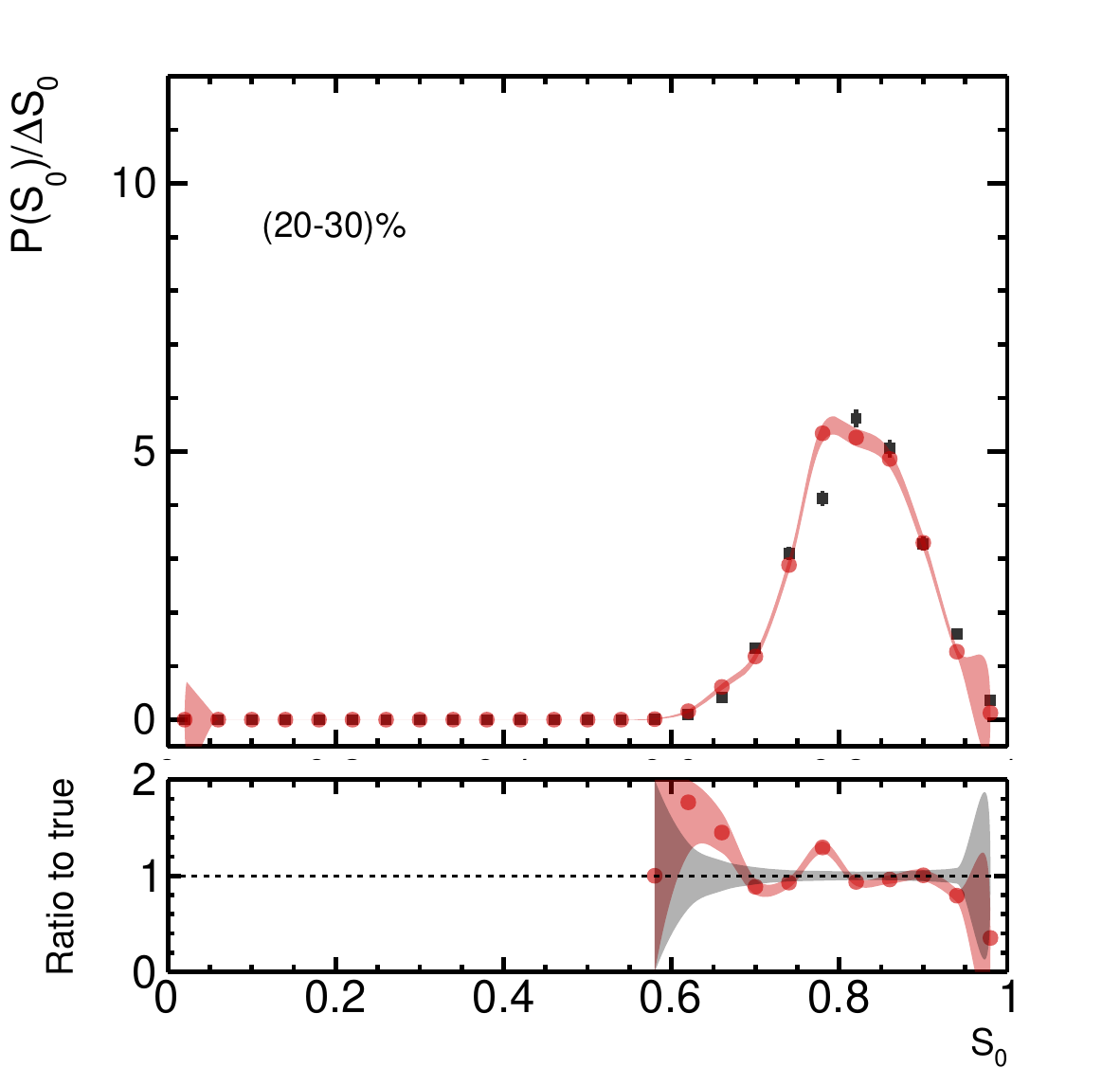}
	\includegraphics[scale=0.38]{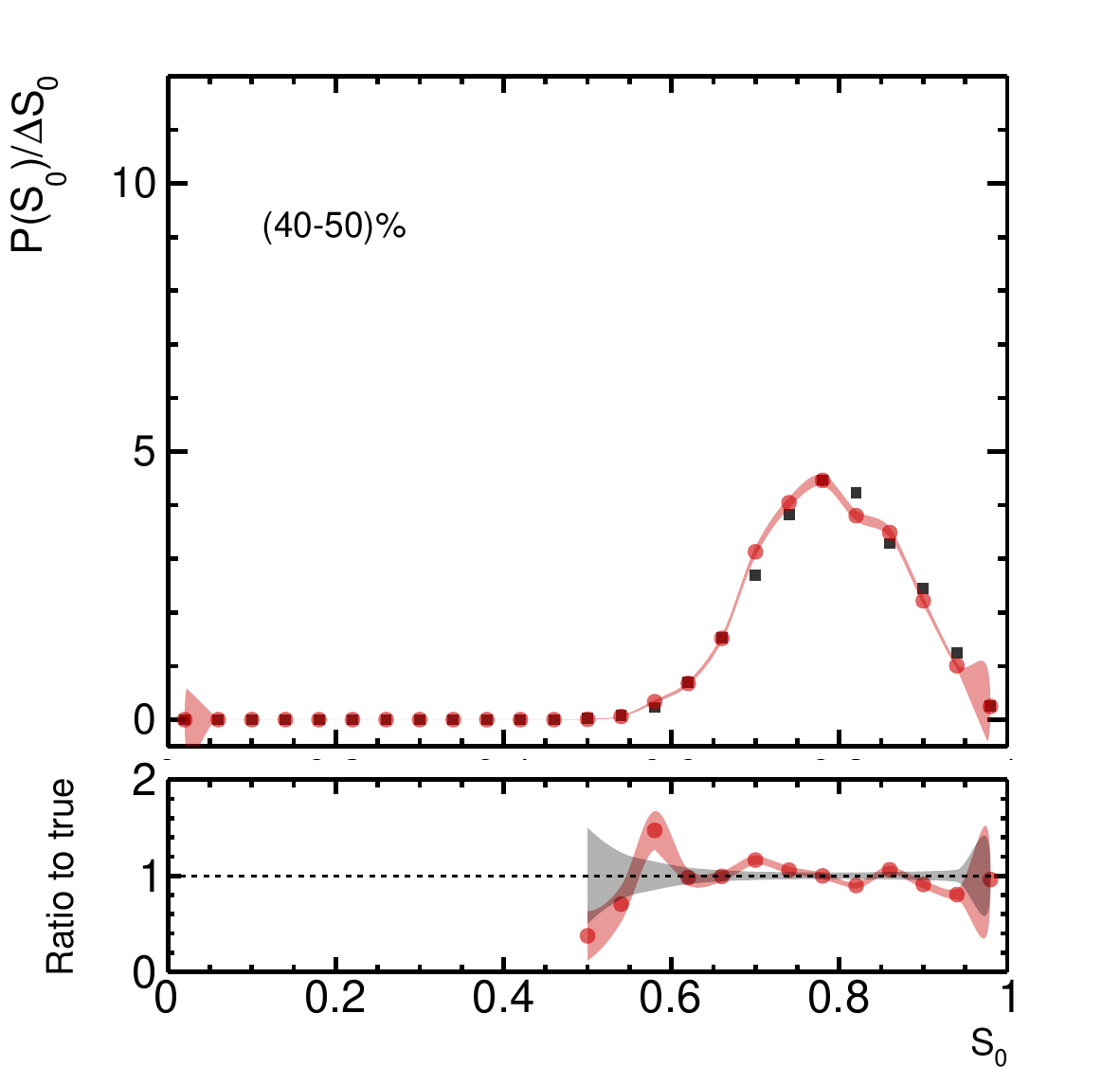}
	\includegraphics[scale=0.38]{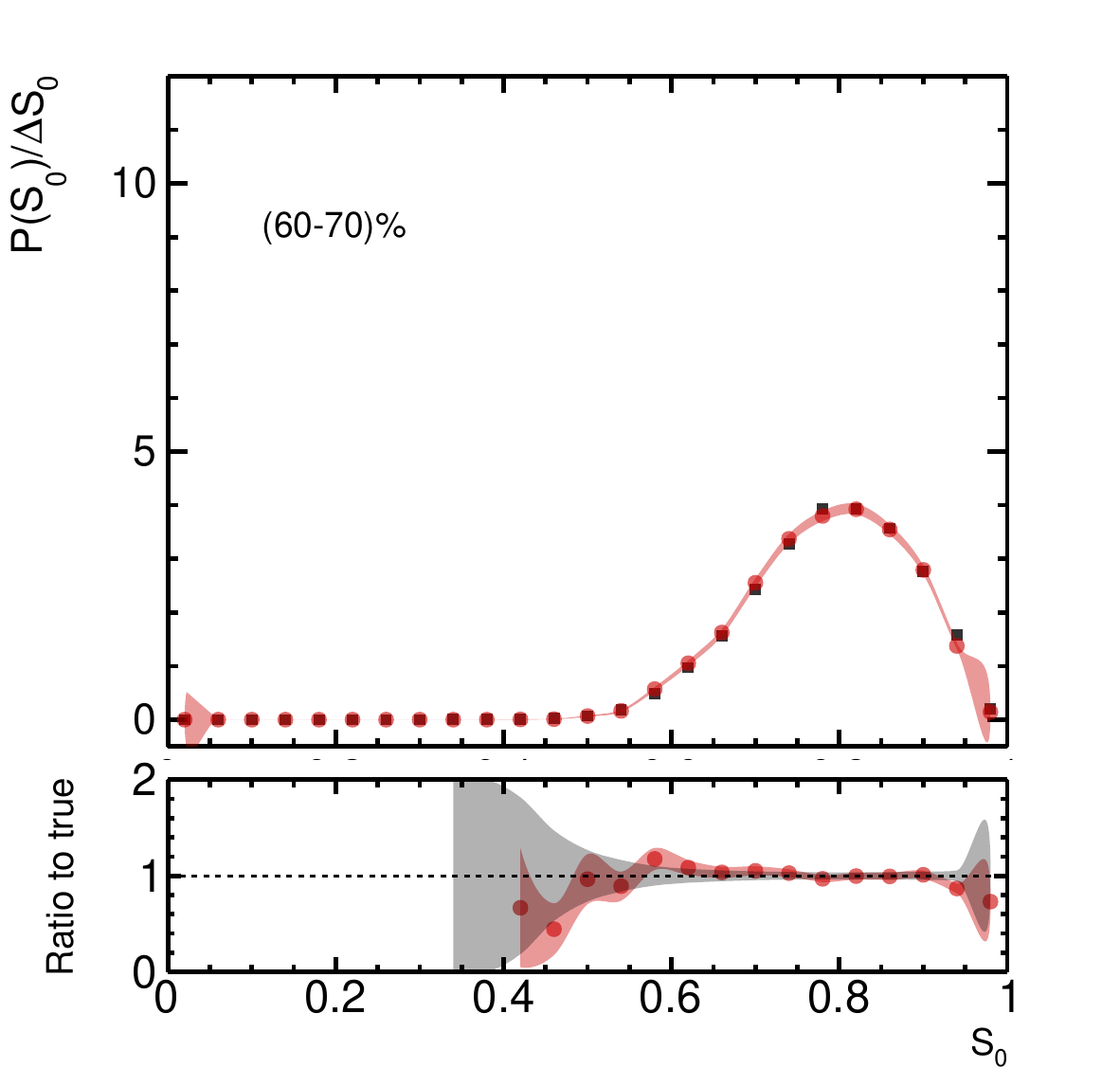}
	\caption[]{(Color Online) Predictions of transverse spherocity distributions for different centrality classes in Pb-Pb collisions using the ML (GBDT) model from minimum bias Pb-Pb collisions at $\sqrt{s_{\rm{NN}}} = 5.02$~TeV in AMPT model. The lower panels show the ratio of the predicted values to the true values. The quadratic sum of the statistical and systematic uncertainties are shown as a red-colored band for the predicted values. The statistical uncertainties in the true values are shown as bars. In the ratio, black-colored band denotes the statistical uncertainties in the true values while the red-colored band denotes the quadratic sum of statistical and systematic uncertainties.}
	\label{Sphero-Cent-AMPT}
\end{figure*}

Bottom panel of Fig.~\ref{CorMatrix} shows the correlation matrix of the input variables and transverse spherocity in Pb-Pb collisions at $\sqrt{s_{\rm{NN}}} = 5.02$~TeV. The numbers show the correlation coefficients ($\rho$) which is obtained from Eq.~\ref{rho}. The values of $\rho$ for intersecting variables in the matrix suggest there are good dependency of the chosen input variables and the transverse spherocity. Based on Eq.~\ref{eq1}, one would naively expect that spherocity would be highly correlated to the charged-particle multiplicity and mean transverse momentum of an event. Thus, we have chosen total charged particle multiplicity, charged particle multiplicity in the transverse region and mean transverse momentum as the input variables for ML prediction. From Fig.~\ref{CorMatrix}, it is found that the transverse spherocity has very high correlation with total charged particle multiplicity and charged particle multiplicity in the transverse region of an event. Although, the correlation with mean transverse momentum is small but it is still significant for a proper prediction of transverse spherocity through ML. To understand the correlation between the input variables and the transverse spherocity we have shown the correlation between each input variable and the predicted value of transverse spherocity in Pb-Pb collisions at $\sqrt{s_{\rm{NN}}} = 5.02$~TeV in Fig.~\ref{CorrPlots-sphero}. One could observe that in the top and middle plots, the high-spherocity region is highly correlated with $\langle dN_{\rm ch}/d\eta \rangle$ and  $\langle N_{\rm ch}^{\rm TS}\rangle$  of an event. We observe that the events with high-spherocity consist of large number of final state charged particles. However, low-spherocity region tends to a back-to-back structure and consequently the correlation between transverse spherocity and charged-particle multiplicity decreases. However, in this region, the $\langle p_{\rm T} \rangle$ plays a bigger role as the transverse momentum of the produced particles are expected to be high. We have also studied the correlation of transverse spherocity with leading-transverse momentum of an event and charged-particle multiplicity in the towards and away region. However, their effects are found to be quite negligible in the ML prediction. Thus, we have only considered the shown input variables in Fig.~\ref{CorrPlots-sphero} for our present study. Let us now move to the predictions of the transverse spherocity and see how they compare with their true values.

Figure~\ref{Sphero-AMPT} shows the predictions for transverse spherocity distribution in Pb-Pb collisions at $\sqrt{s_{\rm{NN}}} = 5.02$~TeV. Here we have also compared the predicted values with the true spherocity distribution obtained from AMPT. One can clearly see that the proposed ML framework with  $\langle dN_{\rm ch}/d\eta \rangle$,  $\langle N_{\rm ch}^{\rm TS}\rangle$ and $\langle p_{\rm T} \rangle$ as the input variables predicts the spherocity distribution accurately in Pb-Pb collisions at $\sqrt{s_{\rm{NN}}} = 5.02$~TeV. However, at low-spherocity regions, we see deviation from the true distribution and this could be due to the fact that in heavy-ion collisions the statistics of having events with back-to-back structure are expected to be quite less compared to events with isotropic in nature. Thus, we believe that this deviation could be due to limited statistics in the low spherocity region, which can also be seen by the black-colored band in the lower panel. In the bottom plot, we have obtained the estimation of spherocity distribution from the input variables in Pb-Pb collisions at $\sqrt{s_{\rm{NN}}} = 2.76$~TeV with the ML training from Pb-Pb collisions at $\sqrt{s_{\rm{NN}}} = 5.02$~TeV. We observe that ML could successfully predict the spherocity distribution at $\sqrt{s_{\rm{NN}}} = 2.76$~TeV in wide spherocity ranges. This suggests that the correlation of spherocity distributions with the input variables are quite similar across LHC energies.

To understand if the proposed algorithm is affected by a particular Monte-Carlo (MC) model, we have used the similar ML algorithm in PYTHIA8 (Angantyr) model in Fig.~\ref{Sphero-PYTHIA}. As evident in Sec.~\ref{section2}, the physics mechanisms in AMPT model and PYTHIA8 (Angantyr) are quite different. However, in Fig.~\ref{Sphero-PYTHIA}, we observe that the predictions for transverse spherocity distribution for Pb-Pb collisions at $\sqrt{s_{\rm{NN}}} = 5.02$~TeV in PYTHIA8 model is quite accurate compared to the true distribution. After we confirm that the proposed ML algorithm does not have any significant bias due to a particular event generation model, we now move to the predictions of spherocity distribution for different centrality classes in Pb-Pb collisions at $\sqrt{s_{\rm{NN}}} = 5.02$~TeV with the ML training with minimum bias simulated data. Figure~\ref{Sphero-Cent-AMPT} shows the predictions of transverse spherocity distributions for (0-10)\%, (20-30)\%, (40-50)\% and (60-70)\% centrality classes in Pb-Pb collisions at $\sqrt{s_{\rm{NN}}} = 5.02$~TeV. Here the used input variables are for specific centrality classes but the ML training is from minimum bias Pb-Pb collisions at $\sqrt{s_{\rm{NN}}} = 5.02$~TeV. Also, the predicted results are compared with true spherocity distribution and it is found that for high-spherocity regions, the prediction is quite consistent with the true distribution (evident in the lower panels).

The obtained results from AMPT are quite interesting and encouraging. In the absence of experimental data, the proposed ML algorithm gives an important tool to obtain the impact parameter and spherocity distributions using the available observables from experiments such as final state charged-particle multiplicity and mean transverse momentum. It would be very interesting to see
how our results compare with the same from experiments. So, it is quite evident that the current study will act as a baseline for future experimental exploration in this direction.

\section{Summary}
\label{section5}
In summary, we implement the ML-based regression technique via BDT to obtain a prediction of impact parameter and transverse spherocity in Pb-Pb collisions at the LHC energies using A Multi-Phase Transport Model (AMPT) model. We obtain the predictions for centrality dependent spherocity distributions from the training of minimum bias simulated data and find that the predictions from BDT based ML technique matches with true simulated data. In the absence of experimental measurements, we propose to implement Machine learning based regression technique to obtain transverse spherocity from the known final state quantities in heavy-ion collisions.

We would like to mention here that the ML-based training with the correlations of input observables using a MC model is quite useful, 
when the MC model describes the input observables as close as possible to the experimental data. This method will be useful to handle the
physics associated with unmeasured quantities in the experiment. In addition, to handle heavy computational problems of central heavy-ion collisions of high-energy experimental data, such a ML-based training using minimum bias data could be used to deal with centrality dependent
behaviour of observables for a given collision energy and colliding species.

\section*{Acknowledgements}
 R.S. acknowledges the financial supports under the CERN Scientific Associateship and the financial grants under DAE-BRNS Project No. 58/14/29/2019-BRNS. The authors would like to acknowledge the usage of resources of the LHC grid computing facility at VECC, Kolkata. S.T. acknowledges the support from INFN postdoctoral fellowship in experimental physics. S.T. also acknowledges the discussions related to machine learning tools with Dr. Antonio Ortiz. A. N. M. thanks the Hungarian National Research, Development and Innovation Office (NKFIH) under the contract numbers OTKA K135515, K123815 and NKFIH 2019-2.1.11-T ET-2019-00078, 2019-2.1.11-T ET-2019-00050 and the Wigner GPU Laboratory.


\end{document}